\documentclass[aps,nofootinbib,twocolumn,superscriptaddress,prx]{revtex4-1}
\usepackage{graphicx,amsmath, amsfonts,bm, color}

\usepackage{xcolor, bm}
\usepackage[hypertexnames=false]{hyperref}
\hypersetup{
   colorlinks,
   linkcolor={blue!50!black},%{red!80!black},
   citecolor={blue!50!black},
   urlcolor={blue!80!black}
}
\usepackage{etoolbox}
\pretocmd{\eqref}{Eq.~}{}{}

\newcommand{\avg}[1]{\left\langle#1\right\rangle}
\newcommand{\MI}{$\mathcal{M} \ $}
\newcommand{\MIEq}{\mathcal{M}}

\DeclareMathOperator{\tr}{Tr}

\begin{document}
\title{Machine-learning Iterative Calculation of Entropy for Physical Systems}

\author{Amit Nir}
\affiliation{The School of Physics and Astronomy, Tel Aviv University}
\affiliation{The Center for Physics and Chemistry of Living Systems, Tel Aviv University}
\author{Eran Sela}
\affiliation{The School of Physics and Astronomy, Tel Aviv University}
\author{Roy Beck}
\affiliation{The School of Physics and Astronomy, Tel Aviv University}
\affiliation{The Center for Physics and Chemistry of Living Systems, Tel Aviv University}
\affiliation{The Center for Nanoscience and Nanotechnology, Tel Aviv University}
\author{Yohai Bar-Sinai} 
\affiliation{The School of Physics and Astronomy, Tel Aviv University}
\affiliation{The Center for Physics and Chemistry of Living Systems, Tel Aviv University}
\affiliation{Google Research, Tel-Aviv}

\begin{abstract}
Characterizing the entropy of a system is a crucial, and often computationally costly, step in understanding its thermodynamics. It plays a key role in the study of phase transitions, pattern formation, protein folding and more. Current methods for entropy estimation suffer either from a high computational cost, lack of generality or inaccuracy, and inability to treat complex, strongly interacting systems. In this paper, we present a novel method, termed MICE, for calculating the entropy by iteratively dividing the system into smaller subsystems and estimating the mutual information between each pair of halves. The estimation is performed with a recently proposed machine learning algorithm which works with arbitrary network architectures that can be chosen to fit the structure and symmetries of the system at hand. We show that our method can calculate the entropy of various systems, both thermal and athermal, with state-of-the-art accuracy. Specifically, we study various classical spin systems, and identify the jamming point of a bidisperse mixture of soft disks. Lastly, we suggest that besides its role in estimating the entropy, the mutual information itself can provide an insightful diagnostic tool in the study of physical systems.
\end{abstract}

\maketitle
Entropy is a fundamental concept of statistical physics whose computation is crucial for a proper description of many phenomena, including phase transitions~\cite{kardar2007statistical,de1993physics,frenkel1999entropy}, pattern formation~\cite{RevModPhys.65.851}, self-assembly~\cite{asor2017crystallization, cho2005self, Donev990}, protein folding~\cite{Avinery,baxa2014loss,brady1997entropy} and many more. In the physical sciences, entropy is typically interpreted as quantifying the amount of disorder of a system, or the level of quantum entanglement. Entropy is also a fundamental concept in other fields of thought -- statistical learning, economy, inference and cryptography, among others~\cite{mackay2003information}. There it is used to quantify the complexity of statistical distributions. Mathematically, entropy is defined as:
\begin{equation}
    S = -k_{\rm{B}} \sum_i p_i \log p_i,
    \label{eq:S_definition}
\end{equation}
where $p_i$ is the probability that the system is in the $i$-th
microstate, and $k_{\rm{B}}$ is the Boltzmann constant. For convenience, in what follows we work with units where $k_{\rm{B}}=1$.

Analytic calculation of the entropy is achievable only for simple, weakly interacting systems. Experimentally, the entropy can be obtained, for example, by measuring the temperature ($T$) dependence of the specific heat down to low temperatures \cite{kittel1998thermal}. Computationally, for all but the simplest systems, a direct calculation of the entropy is computationally infeasible, as it requires computational resources that grow exponentially with system size~\cite{frenkel2013simulations, hansen2014practical}. For example, a classical numerical approach involves integrating the specific heat, which is inferred from energy fluctuations, down to low temperatures~\cite{kittel1998thermal}. This method is computationally costly and can suffer from inaccuracies for systems with numerous ground-states at low $T$. Other methods estimate directly the free energy~\cite{jarzynski1997nonequilibrium}, 
or embrace additional knowledge on the system, for example from experiment, to reduce the entropic contribution to a manageable computational task~\cite{piana2012protein}. 

Recently, we and others have shown that using compression algorithms one can compute, to a good approximation, the entropy of fairly complex systems~\cite{Avinery, Daan, martiniani2019quantifying}. This method is based on  Kolmogorov's theorem that states that the optimal compression of data drawn from a distribution is bounded by the distribution's entropy~\cite{Shannon1948, Kolmogorov1958}. The compression-based methods capitalize on decades of research in computer science, which resulted in fast and efficient compression algorithms, such as the Lempel-Ziv algorithm or variants of it~\cite{LZ} which are widely available. However, these algorithms treat data as a one-dimensional (1D) discrete string, and manipulating higher dimensional data into a 1D structure results in information loss. For example, it was recently demonstrated that compression-based algorithms misestimate the entropy of systems with long-range correlations and fails to capture delicate transitions in complex systems~\cite{Daan}.

Here, we introduce a generic approach which we term \emph{MICE}: Machine-learning Iterative Calculation of Entropy. Our method improves on existing methods in a number of ways: first, it provides state-of-the-art accuracy. Second, it is scalable, in the sense that its computational cost grows logarithmically with system size. Third, it provides estimations of the actual entropy, with physical units, without additive or multiplicative corrections and with no fitting parameters. Fourth, since the underlying computations are performed with artificial neural nets, \emph{MICE} can be naturally applied to various physical systems by adjusting the network architecture, rather than the digital representation of the system (e.g.~flattening high-dimensional systems to one-dimensional byte arrays as in~\cite{Avinery,Daan,martiniani2019quantifying}). Lastly, it can be applied to both discrete and continuous distributions.

Below we test \emph{MICE} on several canonical systems: the Ising model on both square and triangular lattices, the XY model with and without an external magnetic field ($H$) and an athermal system of bi-disperse soft disks in 2D. We show that our approach provides state-of-the-art accuracy, and provides insightful information about the physics as a by-product.

\section*{The method}

\begin{figure}
\centering
\includegraphics[width=0.95\linewidth]{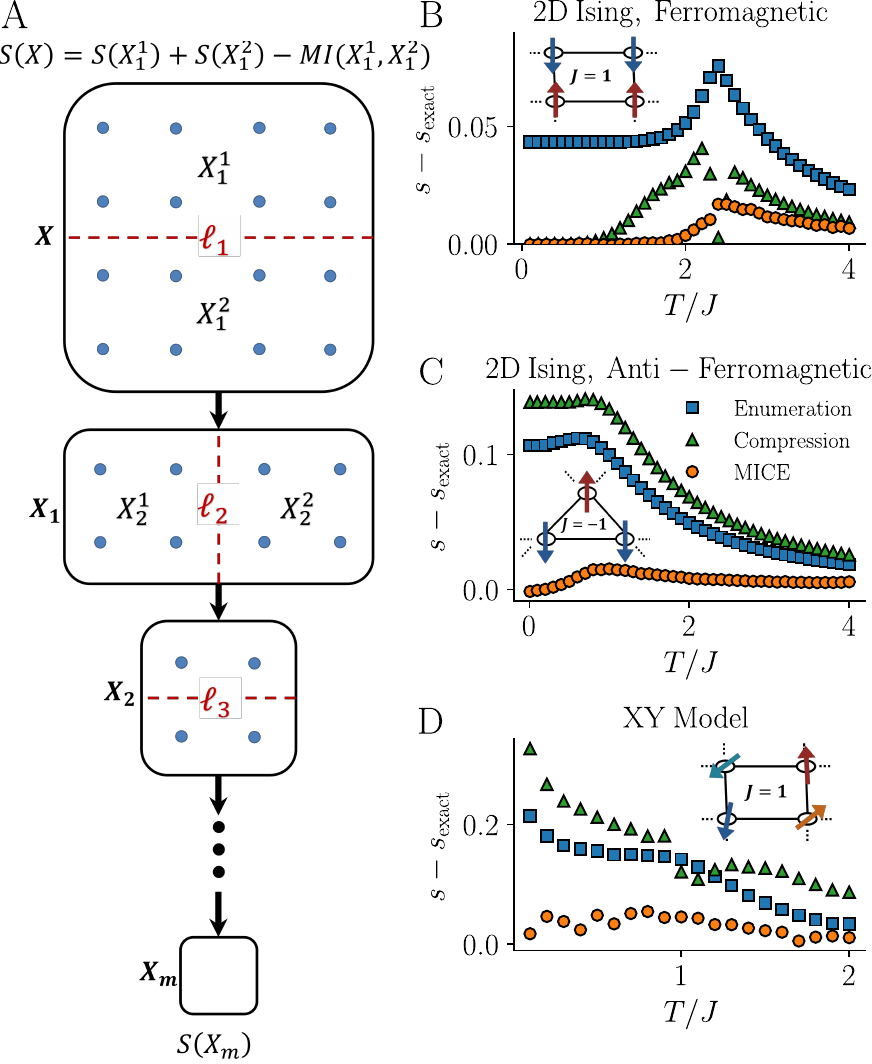}
\caption{(A) Schematic illustration of \emph{MICE}. By dividing into smaller subsystems and calculating the mutual information between them we reconstruct the entropy of the whole system. The entropy of the smallest subsystem is calculated directly by enumeration. Dashed red lines mark the length of interface ($\ell_i$) between two subsystems in the $i$-th iteration. (B-D) The difference between \emph{MICE} estimations of $s$ and known benchmarks. Note that the units are chosen such that $k_B=1$. We present three estimation methods: \emph{MICE}, na\"ive extrapolation from a system of 16 spins (see text) and a compression-based method~\cite{Avinery}. \emph{MICE} shows superior performance in all cases. The three panels show results for (B) ferromagnetic Ising model on a square lattice (C) antiferromagnetic Ising model on a triangular lattice and (D) XY model on a square lattice. In panels B, C we benchmark against known analytical results for infinite systems~\cite{WannierTriangular} and~\cite{Onsager}, respectively. In panel D, we benchmark against the HOTRG calculation of~\cite{HOTRGXY}.}
\label{fig:Figure1}
\end{figure}

\subsection{Entropy and Mutual Information}
In thermodynamics, entropy is considered to be an extensive quantity, i.e.~a quantity that scales linearly with system size. This is only approximately true. In fact, the entropy is strictly \emph{sub-extensive}. The quantity that measures the sub-extensiveness is called mutual information.

To be precise, the mutual information ($\mathcal{M}$) between two random variables $A$, $B$ is defined by the following relation~\cite{mackay2003information}:
\begin{equation}
    S(A, B) = S(A) + S(B) - \MIEq(A,B) \ ,
    \label{entropywithmieqn}
\end{equation}
where $S(A), S(B)$ are the entropies of $A$ and $B$, respectively, and $S(A,B)$ is their joint entropy. It is easy to show that $\MIEq(A,B)$ is strictly non-negative~\cite{mackay2003information}. Therefore, if we think of $A$ and $B$ as two halves of a thermodynamical system, this equation tells us that the entropy of the joint system is smaller than the sum of the entropies of its components. 

\eqref{entropywithmieqn} is the basic relation on which our method relies. It allows calculation of the entropy of a large system by estimating the entropy of each of its halves and the mutual information between them. Since the computational cost of estimating the entropy grows exponentially with the system size, the latter might be a significantly easier problem than the former.

With this in mind, consider a large physical system $X_0$, of volume $V_0$, which we divide to two equal halves. If we deal with translationally invariant systems, as we will assume for the remainder of this work, the two halves are statistically indistinguishable, and we'll denote both of them by $X_1$ (Fig.~\ref{fig:Figure1}A). With this notation, \eqref{entropywithmieqn} takes the form
\begin{equation}
    S(X_0) = 2S(X_{1}) - \MIEq(X_{1}) \ ,
    \label{eq:recursion}
\end{equation}
where $\MIEq(X_k)$ is a shorthand notation for the mutual information between two neighboring subsystems $X_k$. Each of the halves can be further divided into two statistically indistinguishable halves, and this process can be iterated arbitrarily many times. After $m$ iterations, we find that
\begin{equation}
    s(X_0) \equiv \frac{S(X_0)}{V} = s_m  - \frac{1}{2}\sum_{k=1}^m \frac{\MIEq(X_k)}{V_k}  \ ,
    \label{eq:recursion_result}
\end{equation}
where $V_k=2^{-k}V_0$ is the volume (or area in two dimensions) of the $k^{\rm th}$ subsystem, and $s_m\equiv S(X_m)/V_m$ is the specific entropy of the $m^{\rm th}$ subsystem.

\eqref{eq:recursion_result} decomposes the entropy $S$ into contributions from different length scales. At very short scales, the iteration should only be carried out until $X_k$ becomes small enough that its entropy can by directly calculated, either by brute-force enumeration or using other methods. Since $V_k$ decreases exponentially with $k$, the number of needed iterations is logarithmic in the system size. In many cases the actual value of the first term in the right-hand-side of \eqref{eq:recursion_result}, i.e.~the entropy of the smallest subsystem, is an uninteresting additive constant with no physical significance and can be ignored.

In summary, the crux of our method is replacing the problem of evaluating the entropy by that of calculating the mutual information between subsystems of varying sizes, cf.~Fig.~\ref{fig:Figure1}A. It is left to understand how to actually calculate the mutual information, which is the topic of the next section.

\subsection{Estimating the Mutual Information}

Recently, Belghazi et.~al.~proposed a method to calculate the mutual information between high dimensional random variables with neural networks~\cite{mine}. Their idea is simple and elegant: following a theorem by Donsker and Varadhan \cite{donsker1983asymptotic}, the mutual information between two variables, $A$ and $B$, can be expressed as a solution to a maximization problem:
\begin{equation}
\MIEq = \underset{\theta\in\Theta}{\operatorname{sup}} \left[\big\langle \mathcal{F}_\theta(A,B)\big\rangle_{P_{A,B}} - \log\big\langle e^{\mathcal{F}_\theta(A,B)}\big\rangle_{P_{A\times B}}\right].
\label{eq:MI_as_optimization}
\end{equation}
Here, $\mathcal{F}_\Theta :A\times B \rightarrow \mathbb{R}$ is a family of functions parameterized by a vector of parameters $\theta$, $P_{A,B}$ is the joint distribution of $A$ and $B$, and $P_{A\times B}$ is product of their marginal distributions. In our case, since $A$ and $B$ are subsystems of a bigger system, $\avg{\cdot}_{P_{A,B}}$ means averaging over samples of $A$ and $B$ taken from the same sample of the bigger system, while $\avg{\cdot}_{P_{A\times B}}$ means averaging over samples of $A$ and $B$ taken independently. Heuristically, the reason that this representation works is that the mutual information measures how much the joint distribution differs from the product of marginal distributions. In fact, $\MIEq(A,B)$ equals the Kubleck-Leibler divergence between these two distributions~\cite{mackay2003information}. 

While there is much to be said about \eqref{eq:MI_as_optimization}, for the purpose of this work it suffices to note that it reduces the problem of calculating \MI to an optimization problem, which naturally suggests the prospect of using artificial neural networks (ANNs) to parameterize the function $\mathcal{F}_\theta$. This is the core idea of Belghazi et.~al~\cite{mine}, which we adopt. In Machine-Learning language, \eqref{eq:MI_as_optimization}  is taken to be the \emph{loss-function} of the network.

The complete implementation details are given in the supplementary information, Sec.~1. In broader strokes, the process is as follows: first, using standard methods, a sizable dataset of samples of the system is produced. Then, for each size of subsystem pair we generate two datasets: one in which the two subsystems are taken from the same larger sample (the ``joint'' dataset) and another in which each subsystem is sampled independently (the ``product'' dataset). Then, each of the datasets is fed to an ANN, the two averages in \eqref{eq:MI_as_optimization} are calculated, and the weights of the ANN are updated to maximize the loss. This process is repeated until the loss stops improving and \MI saturates. We found exponential moving average useful to reduce noise when estimating \MI over the final training epochs. Finally, \MI is calculated from the trained ANN by averaging \eqref{eq:MI_as_optimization} over a separate dataset, different from the one used to train the network. 

\section*{Results}
To demonstrate the performance and versatility of \emph{MICE} we chose four systems representing different classes of collective behavior: (a) the 2D ferromagnetic Ising model on a square lattice with coupling constant $J=1$, a canonical example of a system with a second order phase transition; (b) the anti-ferromagentic Ising model on a triangular lattice ($J=-1$), a canonical example of a frustrated system with degenerate ground states \cite{landau2014guide}; (c) the continuous XY model on a square lattice, which has a continuous symmetry and features a topological phase-transition \cite{landau2014guide}; (d) lastly, we analyze an athermal system of a bidisperse mixture of elastic particles which undergoes a jamming transition when its density is increased above a certain threshold~\cite{JammingPoint}. For all these systems our method achieves state-of-the-art performance. In addition, in some cases it provides physical insights about the structure and scales of the emergent behavior, as discussed below.   

\subsection*{Spin models}
All three spin models were simulated for a system of $64 \times 64$ spins with periodic boundary conditions. The distribution was sampled using standard, well-established methods: The Ising models were simulated using Metropolis Monte Carlo simulations as in Ref.~\cite{Avinery} and the XY model was simulated using the Wolff algorithm as in Ref. \cite{XyModelWolff} (see supplementary information Sec.~2). 

Lattice systems can naturally be represented as 2D arrays (the triangular lattice can be represented on a square lattice with diagonal interactions \cite{landau2014guide}). This allows the usage of one of the most successful ANN architectures to parameterize $\mathcal{F}$ of \eqref{eq:MI_as_optimization}: feed-forward convolutional nets~\cite{alexnet, lecun2015deeplearning}. We use $1-3$ convolutional layers, each of $8-16$ filters of size $3\times 3$, followed by $2$ fully connected layers, using RELU activation, implemented in PyTorch~\cite{PYTORCH}. Complete details about the hyper-parameters for each model are given in Sec.~1 of the supplementary information. We calculate \MI between subsystems of sizes ranging from a pair of spins to system size. The entropy of a single spin was trivially calculated using brute force enumeration. 

The deviations of our entropy estimations from known results~\cite{Onsager, WannierTriangular, HOTRGXY} are shown in Fig.~\ref{fig:Figure1}B-D. In all three cases we see impressive quantitative agreement, to a fraction of $k_b$, with no fitting parameters. We also benchmark our results against the recently proposed compression-based algorithm~\cite{Avinery}. Relying on highly-optimized code and treating the system as effectively 1D, the compression-based algorithm is obviously much faster, about 1-2 orders of magnitude in terms of run-time. However, while it captures the trend, it offers substantially inferior accuracy in some cases. For example, the low-temperature behavior of the anti-ferromagnetic Ising model, cf.~Fig.~\ref{fig:Figure1}C, is governed by a thermodynamic number of ground states with long-range correlations. There, the error of \emph{MICE} is smaller by an order of magnitude than the compression algorithm method.

It is insightful to compare the performance against another very efficient, albeit na\"ive, estimation of $s$ - calculating $s$ for a small collection of spins by direct enumeration, and neglecting the mutual information (i.e.,~the last term in \eqref{eq:recursion_result}). In other words, this is assuming that $S$ is extensive. This estimation, which we refer to as ``na\"ive extrapolation'', provides only slightly worse accuracy than the compression method, as seen in Fig.~\ref{fig:Figure1}. In all cases, \emph{MICE} provides the most accurate calculation with a maximal error of $0.06k_B$ per spin for all the systems and across all temperatures. In Sec.~3 of the supporting information we also use \emph{MICE} to estimate the heat capacity, showing it outperforms the standard method based on energy fluctuations, since the latter is hard to sample at low temperatures or near a phase transition.

As presented above, our method requires training an ANN for every temperature. This is computationally costly. For example, a single training run for calculating \MI between two $64\times 32$ systems of the ferromagnetic Ising model takes several minutes on a standard personal computer. If we were to generate all points in Fig.~1 in this method, the computation time would reach a day or two. However, drastic improvements in the calculation time can be obtained by leveraging the similarity of the systems between different temperatures. This is done by using the weights ($\Theta$ in \eqref{eq:MI_as_optimization}) that were obtained by training for a given temperature as the initial conditions of the training process of a different temperature or size. This technique is ubiquitous in the field of Machine-Learning, where it is called ``transfer learning''~\cite{TransferLearning}. In our case it reduces the training time by 1-2 orders of magnitudes. For additional information see Sec.~1F of the supplementary information.  

\subsection*{Mutual Information as a probe}
\begin{figure}
\centering
\includegraphics[width=1.\linewidth]{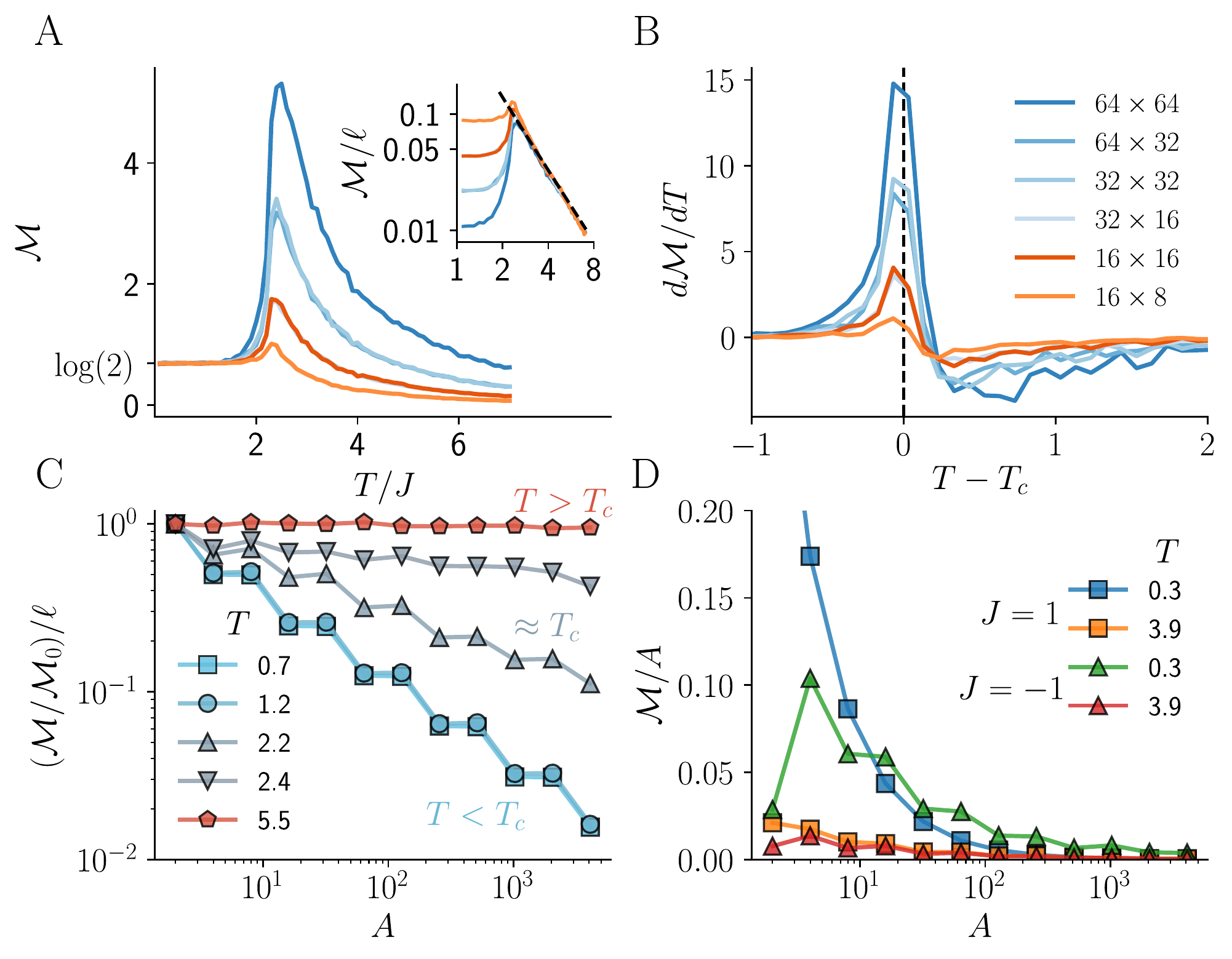}
\caption{Analyzing \MI for the 2D Ising model on a square lattice for different subsystem sizes. (A) \MI complies with two known limits: At low temperatures $\mathcal{M} = \log(2)$. At high temperatures \MI approaches the theoretical value of \eqref{eq:Eran}, as shown in the inset (dashed line). (B) The derivative of the mutual information peaking at the theoretical value $T_c \approx 2.269J$. (Ref. \cite{Onsager1944}). (C) \MI 
normalized by the interface length for varying subsystem sizes (i.e.~number of spins). For visual clarity, all curves are normalized to start at unity at zero area.  (D) \MI per area as function of area for the ferromagnetic Ising model on a square lattice (squares) and the anti-ferromagnetic triangular lattice model (triangles) at various temperatures. \MI decays faster for the ferromagnetic model, as the correlation lengths are much shorter.}
\label{fig:Figure2}
\end{figure}

The main purpose of \emph{MICE} is providing an accurate estimation of $S$. In addition, the byproducts of the calculation, namely the mutual information between systems at different sizes, which is essentially a decomposition of the entropy to contributions from different length scales, can be an interesting observable in its own right. Here we briefly discuss how it captures insightful aspects of the thermodynamics and can be used to assess the accuracy of the \emph{MICE} against known limiting behaviors. In passing we note that the mutual information \emph{between} different scales was shown to be informative in analysis of disordered systems~\cite{Nussinov1, Nussinov2}.

First, we look at \MI between subsystems at various sizes for the ferromagnetic Ising model on a square lattice, plotted in Fig.~\ref{fig:Figure2}. \MI manifestly shows the phase transition~\cite{wilms2011mutual, iaconis2013detecting}. Indeed, $d\MIEq/dT$ peaks\footnote{In second order phase transitions the entropy is continuous but its temperature-derivative (which is proportional to the heat capacity~\cite{kardar2007statistical}) diverges. Since $S$ is a sum over $\mathcal{M}(X_i)$, cf.~\eqref{eq:recursion_result}, we expect $d\mathcal{M}/dT$ to diverge, rather than $\mathcal{M}$.}  exactly at the theoretical infinite-system critical temperature $T_c=2.269J$, cf.~Fig.~\ref{fig:Figure2}B. 

In addition, the accuracy of our calculation can be corroborated against known limits at both high and low temperatures. For $T \ll T_c$, all spins essentially point in the same direction. To be precise, in the low $T$ limit the ground-state entropy of the whole system, or any subsystem, is exactly $\log(2)$. This implies that the mutual information between any two subsystems is also $\log(2)$ which we indeed observe for all subsystem sizes, cf.~Fig.~\ref{fig:Figure2}A.

For $T \gg T_c$, the mutual information between two subsystems can be obtained by a rigorous high-$T$ expansion. The calculation is straightforward but lengthy, and for the sake of brevity its details are given in the Sec.~4A of the supplementary information. However, the result is short and intuitive: the leading order behavior at high $T$ is
\begin{equation}
    \begin{split}
        \MIEq&=\frac{1}{2} \frac{\ell}{ T^2} \ , \qquad \mbox{for Ising model}\\
        \MIEq&=\frac{1}{4} \frac{\ell}{ T^2}\ ,           \qquad \mbox{for XY model}
    \end{split}
    \label{eq:Eran}
\end{equation}
where $\ell$ is the interface size between the subsystems, i.e. the number of spins in one system that directly interact with spins in the other. As seen in Fig.~\ref{fig:Figure2}A (inset), our method shows excellent agreement with this prediction, again with no fitting parameters. In passing we note that \eqref{eq:Eran} is akin to the famous area law in quantum entanglement \cite{wolf2008area}. 

That is, when $T>T_c$ the mutual information per interface length is independent on the system size, as expected. However, for $T<T_c$ the entropy is not extensive, and $\MIEq/\ell$ decays quickly with the size of the subsystem (Fig.~\ref{fig:Figure2}C). This means that the summands in \eqref{eq:recursion_result}, which are \MI normalized by the 2D volume (i.e.~area), decay quickly for large subsystems.  This is visualized in Fig.~\ref{fig:Figure2}D. The figure also shows that in the antiferromagnetic model the summands decay more slowly, which is expected since it features long range correlations.

\begin{figure}
\centering
\includegraphics[width=\linewidth]{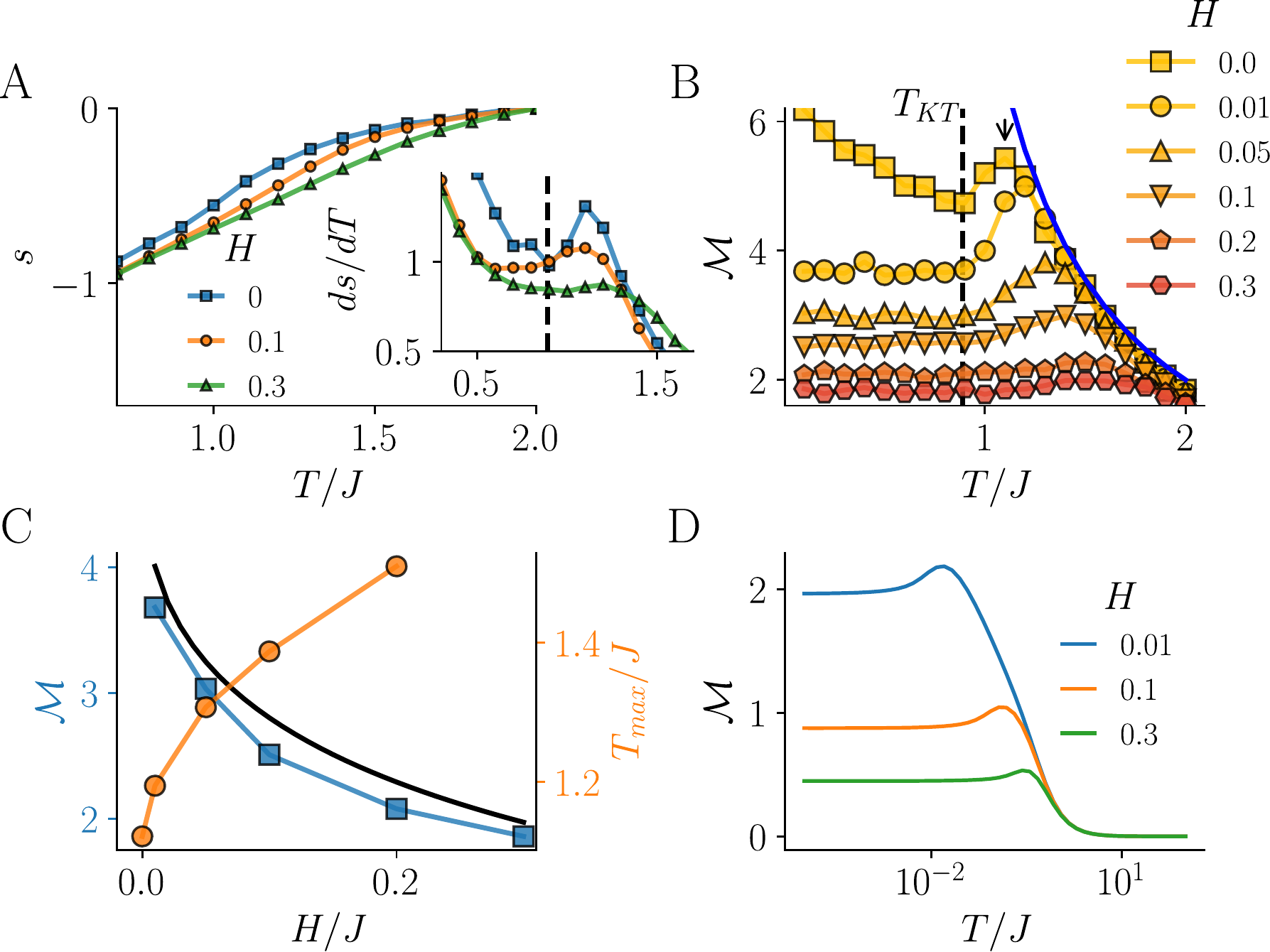}
\caption{Analysis of the XY model under external field ($H$) using \emph{MICE}.
(A) Entropy as a function of temperature for various external fields. Inset shows $ds/dT$, and $T_{KT}$ is marked with a dashed line.
(B) Mutual information between two systems of size  $32 \times 16$ spins,  for varying fields. The arrow marks the peak in $\MIEq(H=0)$ at $T_{\rm max}$. The blue line is the high temperature limit, \eqref{eq:Eran}. 
(C) Two features of the curves at panel B are replotted: The low $T$ plateau value (evaluated at $T=0.1J$), compared to the analytically calculated values at $T \to 0$ in the harmonic approximation, $\MIEq_h$ (black line). $T_{\rm max}$ is plotted in orange circles. 
(D) Exact numerical calculation of $\mathcal{M}$ between two isolated spins for varying $H$, showing qualitatively similar behavior to panel B (though note that the temperature axis is logarithmic, unlike panel B).}
\label{fig:XYResults}
\end{figure}

Next, in Fig.~\ref{fig:XYResults} we examine the entropy and the mutual information in the XY model. At high temperatures \MI decays as described in \eqref{eq:Eran}. Below the critical temperature, the famous Kosterlitz-Thouless transition temperature $T_{KT}=0.8J$, \MI approaches an $T$-independent plateau for $H\ne0$ and diverges logarithmically when $H=0$. This divergence is due to the continuous degeneracy of the XY model, which is lifted in the presence of  an external field. In the transition between these limits, \MI features a pronounced peak, which becomes smaller and shifts to higher temperatures with increasing $H$, cf.~Fig.~\ref{fig:XYResults}C.

This rich behavior of \MI can be understood in simple terms. The high temperature behavior is accurately described by \eqref{eq:Eran}, which is a further corroboration of our method, cf.~Fig.~\ref{fig:XYResults}B. The low temperature behavior can be understood, much like in the case of the Ising model, in terms of collective behavior. For $H\ne 0$ and $T < T_{KT}$ all spins are mostly aligned with the field, even if it is relatively small, because of the broken symmetry. In this case, spins fluctuate mildly around their ground state and a harmonic approximation can be made. Within the harmonic approximation the mutual information, $\MIEq_h$, (the subscript $h$ stands for harmonic) can be obtained analytically in terms of block-determinants of the Hamiltonian, a derivation which is given in detail in Sec.~4B of the supplementary information. The results of this calculation are presented in Fig.~\ref{fig:XYResults}C and show good quantitative agreement.

Lastly, we remark that the generic behavior of \MI -- a $T$-independent plateau at low $T$ followed by a peak and a power-law decay at large $T$ -- is also present in very small systems. In fact, even a system of two spins behaves in a qualitatively similar way, though the transition temperatures between the regimes are quite different due to the collective behavior of the spins, cf.~Fig.~\ref{fig:XYResults}D, and Sec.~5 of the supplementary information.

\section*{A continuous, out of equilibrium system}
One of the main advantages of \emph{MICE} is that it is very versatile in terms of the systems it can operate on. As long as a well-defined distribution exists and samples can be drawn from it, and as long as the system can be digitally represented in a manner compatible with ANNs, \emph{MICE} should be, at least potentially, applicable. In particular, the scheme presented above can be applied to out-of-equilibrium systems, whose entropy calculation is a challenge both technically and conceptually~\cite{Daan, ariel2020inferring,Avinery,jarzynski1997nonequilibrium,martiniani2019quantifying, Daan, PhysRevX.7.021007}. Clearly, the result of \emph{MICE} will be an estimate of the entropy defined in \eqref{eq:S_definition}, which is the information-theoretic definition of entropy. Relating the result to other thermodynamic properties would depend on the details of the system, which is always the case in calculating thermodynamic properties of out-of-equilibrium systems.

Jammed solids are a prominent class of out-of-equilibrium systems whose entropy plays a crucial role in their dynamics~\cite{liu2010jamming}. In these systems the entropy, which stems from steric interactions, is geometric in nature and measures the number of ways the system's constituents can be ordered in space without overlap. When this depends sensitively on the density, jamming occurs. The jamming transition is also important as it is thought that understanding it would guide us in understanding one of the most important open problems in condensed matter physics - the glass transition, which is also intimately related to entropic effects~\cite{liu2010jamming, cavagna2009supercooled, monasson1995structural}.

As a representative example, we study here a bidisperse mixture of soft disks. This system exhibits a jamming transition at high densities~\cite{JammedSystem}. Several works have attempted to identify the jamming transition of this system: using dynamic properties such as the jamming length scale, or the effective viscosity~\cite{JammingAspects}; using static properties such as pair-correlations or fraction of jammed particles~\cite{JammedSystem, JammingAspects}. Recently, Zu and collaborators~\cite{Daan} tried to measure the entropic signature of the jamming transition, and have shown that compression-based methods have failed to do so. The authors of Ref.~\cite{Daan} have generously shared their dataset with us, to test our method on, which we do below.

\begin{figure*}
\centering
\includegraphics[width=1. \linewidth]{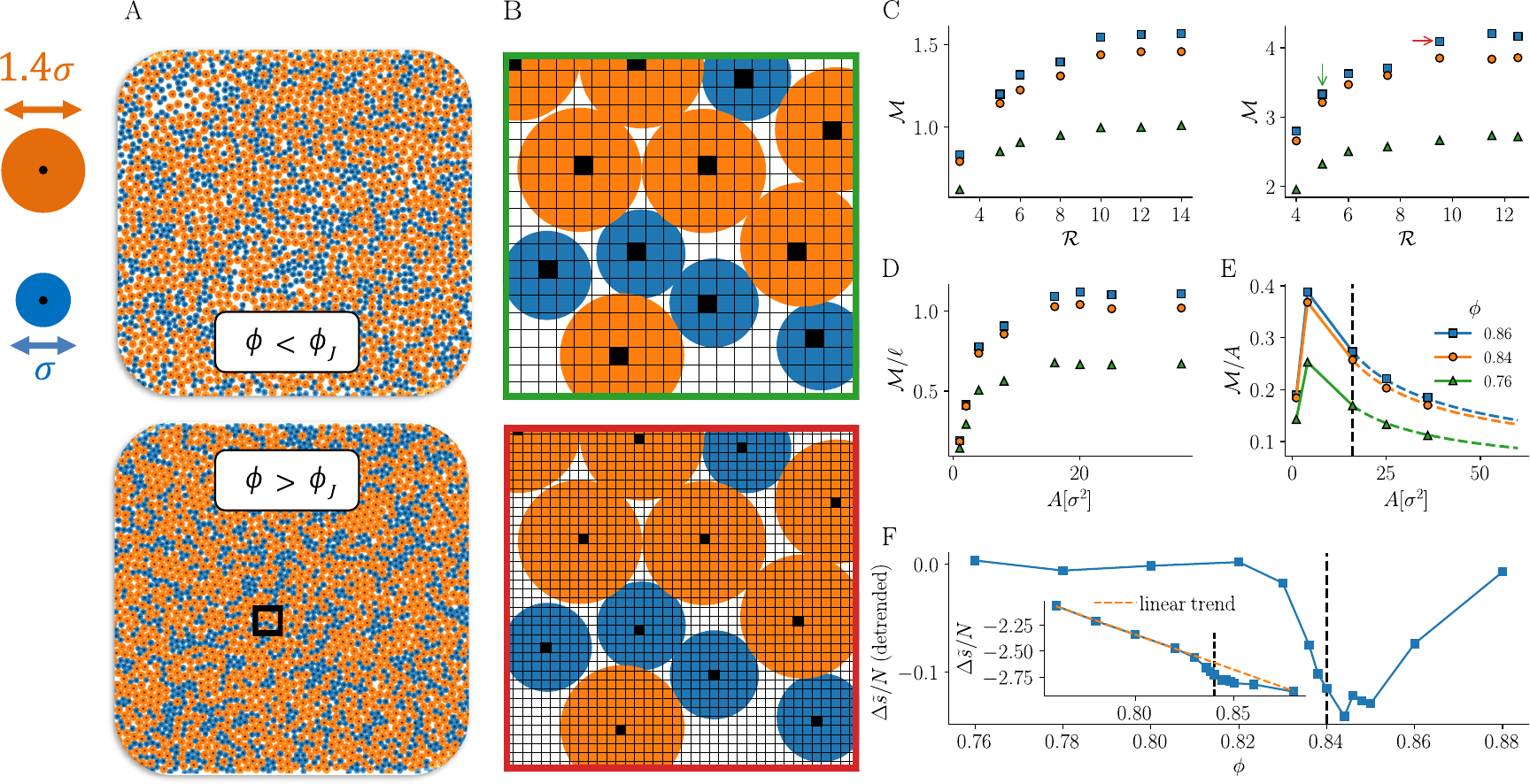}
\caption{(A) Snapshots from the bidisperse mixture simulation below and above the jamming transition density ($\phi_J$).
(B) A  blowup of the marked  region in panel A. We discretized the system (colored  circles) as Boolean 2D images (black and white pixels). The top and bottom panels show a spatial resolution of $\mathcal{R}=5$ and $\mathcal{R}=9.5$, respectively. The pixels are the input to \emph{MICE}. 
(C) The effect of discretizing with various resolutions ($\mathcal{R}$) and various densities. $\mathcal{M}$ between two subsystems of size  $2\sigma \times 1  \sigma$ (left panel), and $4 \sigma \times 2  \sigma$ (right panel). At high resolutions, $\mathcal{M}$ becomes  independent of $\mathcal{R}$. Green and red arrows indicate the resolutions represented in the top and bottom of panel B, respectively. Different markers correspond to different densities, see legend in panel E. 
(D) $\mathcal{M}/\ell$ as function of the area of the subsystem ($A$) at various densities, see legend in panel E. For large enough $\ell$, $\mathcal{M}$ becomes linear in $\ell$. 
(E) $\mathcal{M}/A$ as function $A$ at various densities. $\mathcal{M}$ becomes negligible for large subsystems. The dashed colored lines represent the extrapolation of \eqref{eq:mi_extrapolation}, based on the subsystem at the size represented by the black dashed line. (F) The density dependence of the excess entropy. The inset shows the results of \emph{MICE} (blue), and the linear trend of $\tilde s / N$ at low densities (dashed orange line). For visual clarity, the linear trend in $\phi$ is subtracted in the main panel. The dashed black line represents the theoretical jamming transition point.}
\label{fig:Figure4}
\end{figure*}

The system is an equimolar bidisperse system of disks with one-sided harmonic interactions, cf.~Fig.~\ref{fig:Figure4}A. The simulation is performed in a finite box with periodic boundary conditions. The area density of the particles, $\phi$, is a control parameter which is changed by changing the number of particles, $N$. Further details about the simulation are given in Sec.~6 of the supplementary information. The system is expected to undergo a jamming transition at $\phi_J \approx 0.841$ \cite{JammingPoint, JammingAspects}.

There are a few differences between this system and the spin models discussed above. First, it is not a lattice system with discrete states. Rather, here the state space is continuous, parameterized by the positions of the particles. This requires a careful treatment of the discretization scheme. The choice of discretization scheme, and specifically the spatial resolution of discretization, affects the results in a nontrivial manner. Lastly, in the analysis of the spin models we employed \emph{MICE} on subsystems of all sizes, between 1 spin and the whole system. However, the soft disk systems are so large that doing so will be both impractical and unnecessary (adequate resolution requires $\sim 3\times 10^6$ pixels, as discussed below). Before describing the results, we briefly discuss how these challenges are resolved, since they are common to many physical systems of interest, both in and out of equilibrium.

\paragraph*{Continuous systems (differential entropy)} Since the system is continuous, the summation in \eqref{eq:S_definition} should be replaced by integration: 
\begin{equation}
\tilde S=-\int p(x) \log p(x) dx \ .
\label{eq:diff_s}
\end{equation}
This definition is known as \emph{differential entropy}.
Note that $\log p(x)$ is ill defined since it depends on the choice of units of $x$ in a non-multiplicative manner.

This non-multiplicative component, which depends logarithmically on the length unit, is fundamentally related to the fact that the digital representation of the system is discrete and thus the differential entropy of \eqref{eq:diff_s} differs from the discrete entropy of \eqref{eq:S_definition} by a factor that diverges logarithmically with the resolution of the discretization. This is derived in detail in Sec.~7 of the supplementary information.

Moreover, we also show there that, quite conveniently, the representation of $S$ in terms of \eqref{eq:recursion_result} offers a well defined way to remove this divergence. While $\tilde S$ of a continuous system depends logarithmically on the resolution, $\mathcal{M}$ becomes independent of it in the limit of very fine resolution. In fact, the necessary resolution is such that no physically relevant information is lost by the discretization, i.e.~when all continuous configurations that map to the same discrete representation are equiprobable.

Therefore, when we estimate $S$ according to \eqref{eq:recursion_result} we can avoid the logarithmic divergence simply by omitting the first term in the right-hand-side. That is, in what follows we do not present $\tilde s$ but rather
\begin{equation}
    \Delta \tilde s \equiv \tilde s - \frac{S(X_m)}{V_m} = -\sum_{k=1}^m \frac{\MIEq(X_k)}{2V_k}
    \label{eq:S_without_smallest_scale}
\end{equation}

As a side note, we remark that the omitted term, $S(X_m)/V_m$, is simply the entropy density of the smallest subsystem. It corresponds to the entropy of an ``ideal gas`` composed of copies of the smallest subsystem. Subtracting the entropy of an ideal gas is common in entropy calculations of thermodynamic systems~\cite{Daan, ariel2020inferring}. The result of the subtraction is commonly referred to as ``excess entropy''.

\paragraph*{Discretization} Since convolutional ANNs show state-of-the-art capabilities in extracting information from images, we discretize phase space by mapping a state of the system to a 2D image, whose pixels are black if they contain a center of a particle\footnote{Technically, pixels are black if they contain a center of \emph{one or more} particles, though this never happens in the resolutions we work with.}, see Fig.~\ref{fig:Figure4}B. The spatial resolution of the image is a hyper-parameter of our method. We measure the resolution with the dimensionless number $\mathcal{R}=\sigma/p$, where $p$ is the spatial extent of a pixel and $\sigma$ is the diameter of the smaller disk. Based on the discussion above, we expect the estimation of \MI to converge to a constant value when $\mathcal{R}$ increased. This is indeed the case, as demonstrated in Fig.~\ref{fig:Figure4}C. In what follows, we use a $\mathcal{R}=10$, for which \MI is converged. We note that in terms of resources, the computational cost of discretizating the system is negligible compared to simulating the system or training the ANN. In addition, as shown below, the ANN does not have to be applied on the whole system, so a fine discretization does not lead to a memory bottleneck, at least not in 2D. 

\paragraph*{Extrapolating the mutual information}
The resolution required for convergence necessitates $\sim10^6$ pixels to discretize the whole system. Feeding such a large image to an ANN might be possible, but requires unreasonable computational resources for the task at hand. Luckily, this is not necessary.

As discussed above, for large enough subsystems, that is, scales much larger than the longest correlation length of the system, we expect \MI to grow linearly with the interface length, cf.~Fig.~\ref{fig:Figure2}C. In precise terms, we expect \begin{equation}
    \MIEq(X_k) = \frac{\ell_k}{\ell_n}\mathcal{M}(X_n)\ .
    \label{eq:mi_extrapolation}
\end{equation}
If we assume this is obeyed for all systems larger than $X_k$,  this relation can be used to replace the summands in \eqref{eq:recursion_result}, and the summation can be done analytically without calculations on subsystems larger than $X_k$. Fig.~\ref{fig:Figure4}D shows that this happens for subsystems of length $\sim4\sigma$. In Fig.~\ref{fig:Figure4}E we show that \eqref{eq:mi_extrapolation}, based on the values of \MI for this size, quantitatively reproduces the values of the summands of \eqref{eq:recursion_result} for sizes larger than $4\sigma$, i.e.~a 2D volume of $A=16\sigma^2$.

\paragraph*{Results} 
We are now in position to calculate the entropy of the whole system for various densities. Assuming that \eqref{eq:mi_extrapolation} is satisfied for $n>m$, \eqref{eq:recursion_result} can be analytically summed, yielding (see Sec.~8 of the supporting information):
\begin{equation}
s=s(x_m) - 2\frac{\MIEq(X_m)}{V_m}\ .
\label{eq:eq5}
\end{equation}

The inset of Fig. \ref{fig:Figure4}F shows $\Delta \tilde s / N$ as function of $\phi$. It is seen that at low densities $\Delta \tilde s$ depends roughly linearly on the density (dashed orange line). To emphasize the phase transition, in the main panel we plot the same data with this linear trend subtracted. The change in the behavior of $\Delta \tilde s$ around the expected jamming point is evident. Importantly, we remind the reader that compression-based entropy estimations were less successful in showing this transition (see Sec.~3.5 of~\cite{Daan}). A more detailed comparison with the results of ~\cite{Daan} is given in Sec.~9 of the supplementary information.

\section*{Discussion and Conclusion}
Machine learning algorithms in general, and neural networks in particular, offer an effective tool to identify patterns in high dimensional data with complex correlation structure. We have shown that these capabilities can be leveraged to tackle another important challenge -- computing the entropy of physical systems. 

The crux of the method is mapping the problem of entropy calculation to an iterative process of mutual information estimation. By doing so we were able to estimate the entropy of canonical statistical physics problems, both discrete and continuous, both in and out of equilibrium, outperforming compression-based entropy estimation methods. Lastly, we demonstrated that \emph{MICE} naturally allows to decompose the entropy into contributions from different scales, providing an insightful diagnostic for the thermodynamics of physical systems.

We surmise that \emph{MICE} could be a promising tool for the study of many important systems, such as the configurational entropy of amorphous solids~\cite{bouchbinder2007athermal}, the entropy crisis of glassy systems~\cite{cavagna2009supercooled}, entropy of active matter~\cite{PhysRevX.7.021007}, and more. The main limit of the proposed method would depend on the minimal system size for which \eqref{eq:mi_extrapolation} applies, which determines the largest input for which an ANN should be trained. This is the dominant factor in the computational cost of our method.  In addition, we believe that with adequate modifications \emph{MICE} could be used on quantum systems, for which the mutual information is fundamentally related to entanglement of quantum states~\cite{amico2008entanglement}. A relevant direction could be the extraction of entropy from quantum Monte Carlo simulations. These directions will be explored in future works.

\begin{acknowledgments}
We thank Daan Frenkel, Mengjie Zu and Arunkumar Bupathy for fruitful discussions and for generously sharing their data and code. In addition we thank Yuval Binyamini, Yakov Kantor, Haim Diamant, Gil Ariel and  Amit Moscovich-Eiger for fruitful discussions. We acknowledges support by the Israel Science Foundation (550/15, 154/19), the United States–Israel Binational Science Foundation (201696), and ARO (W911NF-20-1-0013). YBS also thanks his mother.
\end{acknowledgments}
\bibliography{../citations}

\vspace{5cm}

%%%%%%%%%%%%%%%%%%%%%%%%%%%%%%%%%%%%%%%%%%%%%%%%%%%%%%%%%%%%%%%%%%%%%%%%%%%%%%%%%
%%%%%%%%%%%%%%%%%%%%%% these lines of code handle the concatenation %%%%%%%%%%%%%
%%%%%%%%%%%%%%%%%%%%%%%%%%%%%%%%%%%%%%%%%%%%%%%%%%%%%%%%%%%%%%%%%%%%%%%%%%%%%%%%%
\setcounter{equation}{0}
\setcounter{figure}{0}
\setcounter{section}{0}
\setcounter{table}{0}
\setcounter{page}{1}
\makeatletter
\renewcommand{\theequation}{S\arabic{equation}}
\renewcommand{\thefigure}{S\arabic{figure}}
\renewcommand{\thesection}{S\arabic{section}}
\renewcommand*{\thepage}{S\arabic{page}}
% \renewcommand{\bibnumfmt}[1]{[S#1]}
% \renewcommand{\citenumfont}[1]{S#1}
%%%%%%%%%%%%%%%%%%%%%%%%%%%%%%%%%%%%%%%%%%%%%%%%%%%%%%%%%%%%%%%%%%%%%%%%%%%%%%%%%
%%%%%%%%%%%%%%%%%%%%%% these lines of code handle the concatenation %%%%%%%%%%%%%
%%%%%%%%%%%%%%%%%%%%%%%%%%%%%%%%%%%%%%%%%%%%%%%%%%%%%%%%%%%%%%%%%%%%%%%%%%%%%%%%%

\onecolumngrid
\pagebreak
%\vspace{0.25cm}
\begin{center}
	\textbf{\large Supplemental Materials}
\end{center}
\vspace{1cm}
\twocolumngrid
\renewcommand{\MI}{\mathcal{M}}

\section{\emph{MICE} implementation details}
\label{Appendix:Network}

\subsection{Data preprocessing and augmentation}
\label{augmentation}
Input features were normalized between values -1 and  1. For the soft disk system, this means that empty pixels are set to $-1$ and pixels which contain a particle center are set to $1$.
Since all our systems are symmetric under reflections, we performed data augmentation by reflecting both vertically and horizontally. In the data of the XY model without an external field, a global random phase was also used for data augmentation. In addition, due to translational symmetry one can sample subsystems anywhere within the larger system. Combining all these, a single snapshot of 64x64 spins can generate about 15,000 training samples.

\subsection{Network Architecture}
Our method was implemented using the PyTorch library \cite{PYTORCH}. For subsystems of input size larger than $32 \times 32$ we used three convolutional layers with $16$ filters of size $3 \times 3$ each and a rectified linear unit (ReLU) activation. For smaller subsystems, we use only  two convolutional layers. For subsystems of size $4 \times 4$ or smaller, only one convolutional layer is used. The convolutional layers are followed by two fully connected layers, with $\frac{k}{2}$ and 1 output neurons, respectively, where $k$ is the number of output neurons in the last convolutional layer. The batch size for training was 128. 

\begin{figure*}
\centering
\includegraphics[width=.85 \linewidth]{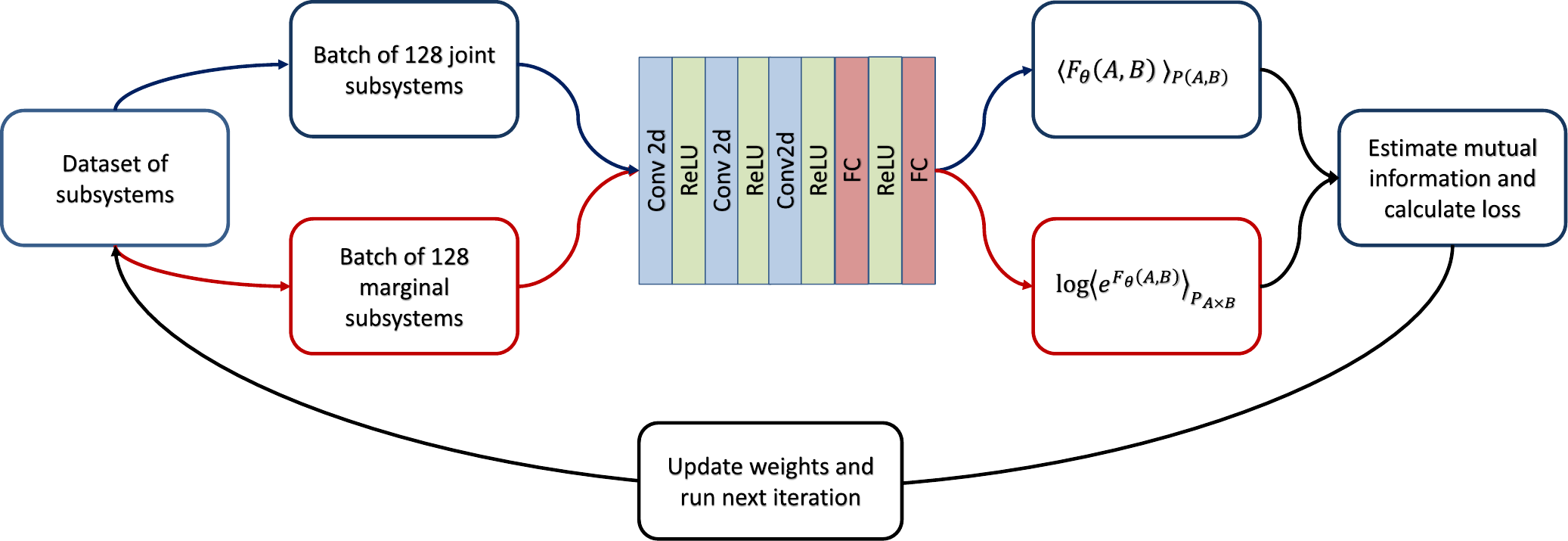}
\caption{The flow of \emph{MICE}. The simulations are used to generate a marginal dataset and a joint dataset (see main text for definition) dataset. The specific architechture of the ANN shown here was used for subsystem pars larger than $32 \times 32$. Smaller subsystems used $1-2$ convolutional layers, as detailed in Sec.~1B.}
\label{fig:NetworkDem}
\end{figure*}

\subsection{Noise Reduction}
\begin{figure}
\centering
\includegraphics[width=\linewidth]{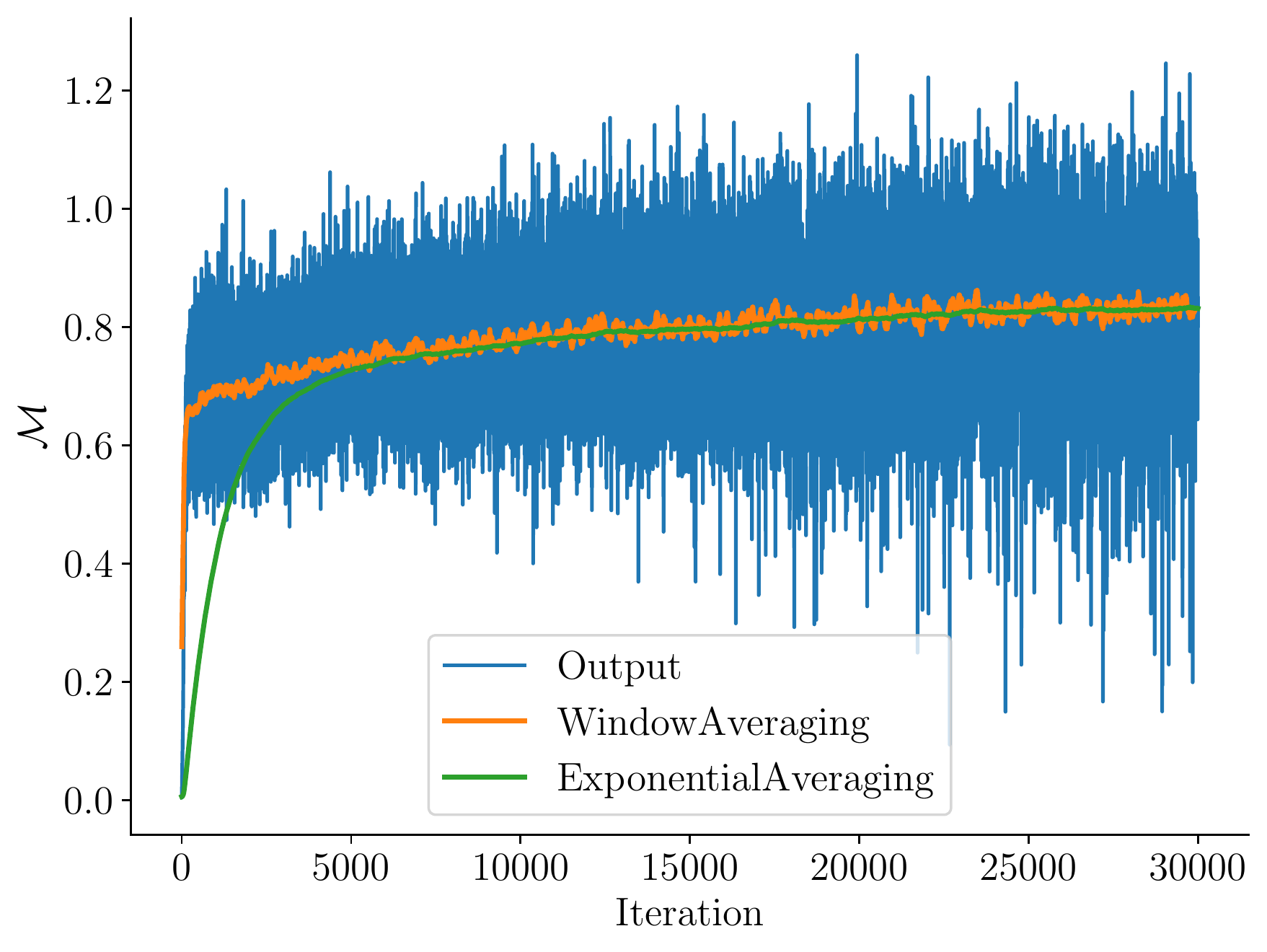}
\caption{Noise reduction. The raw output of the network (blue) and an exponential average with $\alpha=10^{-3}$ (green) are  shown during a typical training loop. In addition, we demonstrate another noise reduction method, used by the original authors of \cite{mine}, a moving average with a window size of 100 iterations (orange).}
\label{fig:Averaging}
\end{figure}

The output of the neural network (ANN) is averaged over the marginal and joint distributions to give a bound on the mutual information (see Eq.~(5) of the main text). As the network learning process progresses, the bound becomes tighter. However, at each iteration the averaging is performed over a small batch of $128$ samples. Therefore, the network's output is extremely noisy. To smooth the results we use a moving exponential average:
\begin{equation}
    \avg{\MI}_{i+1} = \avg{\MI}_{i} + \alpha \Big(\MI_{i+1} - \avg{\MI}_{i} \Big).
    \label{eq:ExpAverage}
\end{equation}
where $\MI_j$ is the output of the network after $j$ optimization iterations, and $\avg{\MI}_{i}$ is our averaged estimation after $i$ iterations, see Fig.~\ref{fig:Averaging}. Throughout the manuscript we used the exponential averaging with $\alpha =10^{-3}$.

\subsection{Validation}
For estimating $\MI$ we implemented the standard scehme of using a validation set. Two independent datasets with ratio of 80-20 were created before training. The network was trained over the large (training dataset), and the training phase was terminated when the $\MI$ estimation on the training set stopped increasing. $\MI$ was estimated over the independent validation set as well, and this value was used for subsequent calculations. By comparing the estimation of $\MI$ over the training and validation sets, one can verify that the network did not overfit the data.

\subsection{Dataset size}
For the spin models we used a dataset of 5000 samples of a $64\times64$ system. An exception is the XY model with an external field where we used 2000 simulations. For the soft disk system we used a set of 100 simulations. In general, for the systems considered in the manuscript we typically needed about $10^4-10^5$ samples (obtained from the the $\sim 10^3$ actual samples by data augmentation, see \ref{augmentation} above) to achieve reasonable convergence.

\subsection{Transfer Learning}
\label{Appendix:TransferLearning}

\begin{figure*}
\centering
\includegraphics[width=\linewidth]{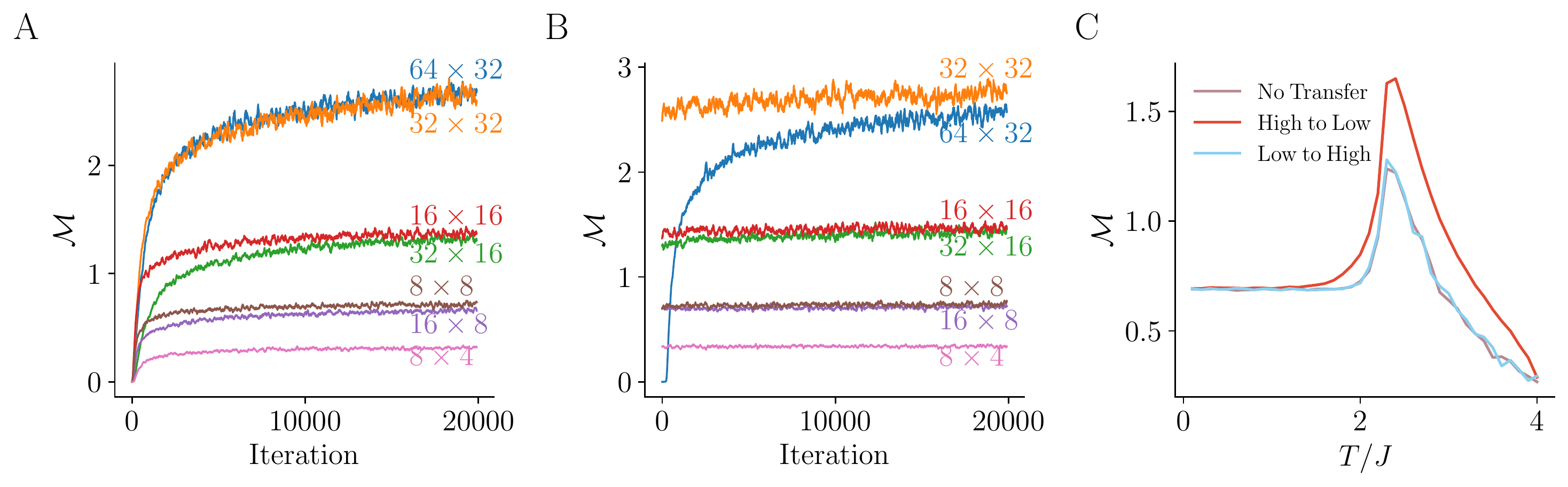}
\caption{Effect of transfer learning. (A)-(B) Learning process as function of iteration for various subsystem sizes. (A) Without transfer learning (i.e.~random initial weights for each ANN). (B) With transfer learning from one subsystem size to another. $\MI$ plateaus at the same level with or without transfer learning, but the number of iterations needed to reach the plateau changes drastically. (C) $\MI$ as function of temperature for $16 \times 16$ subsystem of the 2d ferromagnetic Ising model. Adding transfer learning from high to low temperature improves the results dramatically while transfer learning in the opposite direction is not effective. All trainings were done for $3000$ iterations at every temperature.}
\label{fig:transferNoTransfer}
\end{figure*}

When initiating the network weights at random the resulting estimation of $\MI$ is roughly zero. During training it increases until a plateau is reached. For our choice of hyperparameters this can take a few thousand training iterations, cf.~Fig.~\ref{fig:transferNoTransfer}. This process can be expedited if the network is not initialized at a random initial condition but instead the weights of a network that was trained for a different system are used, a technique called ``Transfer Learning''

This can be done in a number of ways - e.g.~transfer learning across temperatures or the sizes of the subsystem. In the main text we only used transfer learning across different temperatures. In Fig.~\ref{fig:transferNoTransfer} we show the result of training with and without transfer learning, which can reduce training time by 1-2 orders of magnitude. We note that transfer learning works better when we first train on high $T$ and then transfer  to lower $T$, similar to simulated annealing strategy in optimization.

We note that transfer learning across subsystem size is slightly more tricky since the input size to the ANN is different. One simple-minded way to overcome this is to pad the smaller subsystems  with zeros, which gives reasonable results, cf.~Fig.~\ref{fig:transferNoTransfer}B. This is an interesting direction for future research, which we did not further explore. Transfer learning across subsystem size was not used in the main text.

%%%%%%%%%%%%%%%%%%%%%%%%%%%%%%%%%%%%%%%%%%%%%%%%%%%%%%%%%%%%%%%%%%%%%%%%%%%%
\section{Spin Model Simulations}
Sampling the distribution of the Ising systems was preformed using standard Monte-Carlo sampling.

Sampling the distribution of the XY simulation was performed using the Wolff algorithm implemented in the \texttt{c++} library provided in Ref.~\cite{XyModelWolff}. To generate uncorrelated samples the mean cluster size at each temperature, $c$, was calculated and the simulation was sampled every $2/c$ steps. That is, each spin was flipped twice on average between two subsequent samples at all temperatures.

%%%%%%%%%%%%%%%%%%%%%%%%%%%%%%%%%%%%%%%%%%%%%%%%%%%%%%%%%%%%%%%%%%%%%%%%%%%%
\section{Specific Heat Estimation Using \emph{MICE}}

\begin{figure}
\centering
\includegraphics[width=\linewidth]{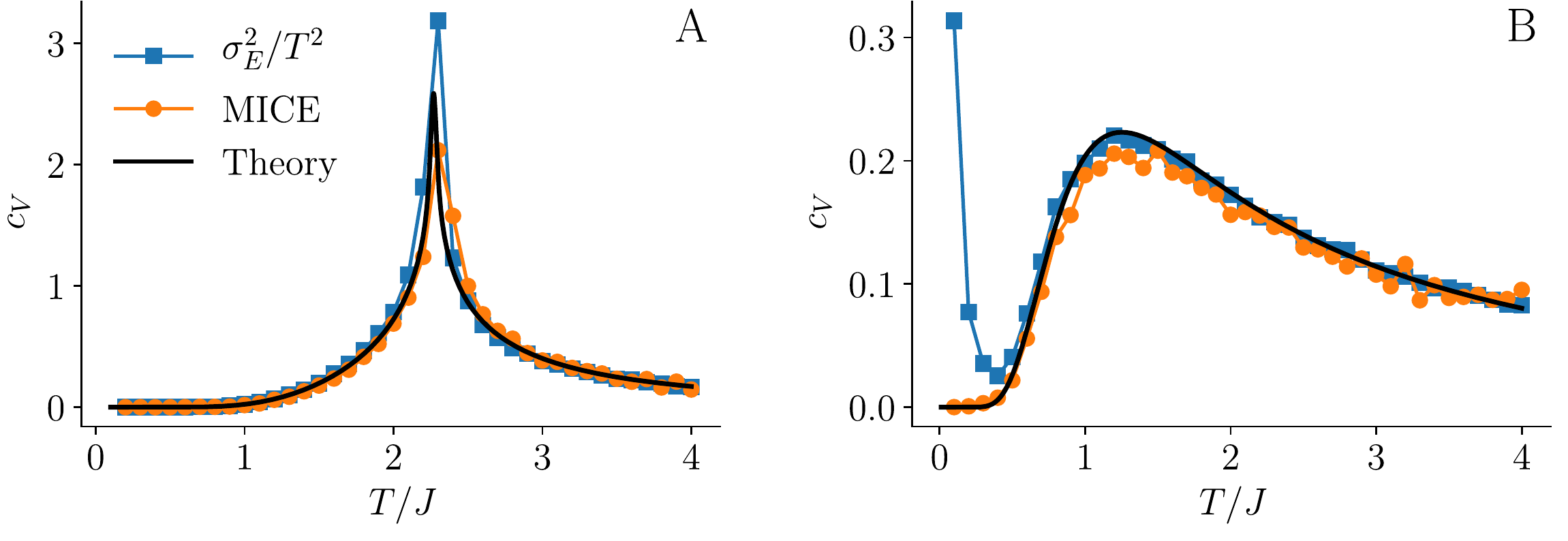}
\caption{Estimating $c_V$ using energy fluctuation estimation,  (\eqref{eq:cv_simulation}, blue), and \emph{MICE} (\eqref{eq:cv_S}, orange), compared to the theoretical value (black). (A) 2D Ising model (B) 2D anti-ferromagnetic Ising model.}
\label{fig:cv_comparison}
\end{figure}
A standard method of estimating the entropy of thermodynamic systems is to integrate the specific heat from low temperatures. This method relies on the relations 
\begin{align}
    c_V &= T\frac{dS}{dT} \ , \qquad \mbox{and}\label{eq:cv_S}\\
    c_V &= \frac{\langle E^2 \rangle - \langle E \rangle ^2}{T^2} \ ,
    \label{eq:cv_simulation}
\end{align}  
where $c_V$ is the heat capacity, $E$ is the energy and $\langle \cdot \rangle$ denotes thermal averaging. $S(T)$ can be calculated using \eqref{eq:cv_simulation} and integrating the energy fluctuations from zero temperature to $T$. 

Alternatively, one take the inverse direction: using the estimation of $S$, as calculated by \emph{MICE}, together with \eqref{eq:cv_S} to estimate $c_V$. In Fig.~\ref{fig:cv_comparison} we compare this estimation of $c_V$ (orange) to the estimation of $c_V$ using energy fluctuations (\eqref{eq:cv_simulation}, blue). It is evident that the energy fluctuations overestimate $c_V$ in the 2D ferromagnetic Ising model near the phase transition, and at low temperatures in the anti-ferromagnetic triangular lattice, which features high degeneracy of low energy states.

%%%%%%%%%%%%%%%%%%%%%%%%%%%%%%%%%%%%%%%%%%%%%%%%%%%%%%%%%%%%%%%%%%%%%%%%%%%%
\section{Analytic Calculation of $\MI$ at high and low temperature limit for spin models}
\subsection{High temperature}
Here we derive Eq.~(6) of the main text by a rigorous high-$T$ expansion of the partition function and marginal probabilities. Physically this expansion relies on the fact that at high temperatures correlations become local. At high temperature we explicitly obtain the area law, $\MI(A,B) \propto \ell$, stating that $\MI$ is proportional to the area $\ell$ (or length in two dimensions) of the interface between regions $A$ and $B$, rather their volume.

The mutual information between subsystems $A$ and $B$, whose union is $A \cup B = X$, is defined as: 
\begin{equation}
\label{eq:MI}
\MI(A,B)=S(A)+S(B)-S(X),    
\end{equation}
where the entropy of a subsystem $A$ is given in terms of the marginal probability:  
\begin{equation}
    S(A) = - \sum_\alpha P_A(\alpha) \log P_A(\alpha).
\end{equation}
Here, $\alpha$ labels microstates of $A$. For the spins models, the microstates are given in terms of the configurations of spins $z_a, a \in A$. We assume that the entire system $X$ under consideration is described by an equilibrium distribution:
\begin{align}
    P_{X} (\bm z) &=\frac{e^{-\beta E(\bm z  )}}{Z}\ , &
    Z &= \sum_{\{z_i = \pm 1 \} } e^{-\beta E(\bm z)}\ .
\end{align}
Here and in what follows boldface letters (e.g. $\bm z$) denote vectors. The marginal distribution of subsystem $A$ is obtained by tracing out the spins in its complement, $P_A  = \tr_{B} P_{X}$.

We proceed by an explicit evaluation of $\MI$ at high temperature for the Ising model:
\begin{align}
    E_{\hbox{Ising}}(\bm z)&=-J \sum_{ \langle i , j  \rangle}  z_i z_j - H \sum_i  z_i\ , &    z_i &= \pm 1 \ .
\end{align}
The expansion of the partition function in powers of $\beta$ up to second order is
%\begin{widetext}

\begin{align}
Z &= \sum_{\{z_i = \pm 1 \} } \left( 1-\beta E(\bm z)  + \frac{1}{2} \beta^2 E(\bm z)^2+ \dots \right)\nonumber\\
&=2^N + \frac{1}{2} \beta^2 \left[ J^2 \left(\sum_{\langle i , j \rangle} 1\right)2^N +H^2 \left(\sum_{i} 1\right)2^N \right] +\mathcal{O}\left(\beta^3\right)\nonumber \\
&=2^N \left[1 +\frac{1}{2} \beta^2  \left(J^2 N_{\rm{links}}+H^2 N  \right) \right] +\mathcal{O}\left(\beta^3\right),    
\label{Zexpand}
\end{align}
where $\sum_{\langle i , j \rangle} 1= N_{\rm{links}}$ is the total number of links and $\sum_{i} 1=N$ is the number of sites. In what follows we omit the external field ($H$) for clarity and conciseness of presentation, and only mention its effect in the end result. 

Next, we perform a high temperature expansion up to order $\beta^2$ of the marginal probability
\begin{align}
\label{PX} &P_A(\bm z_A)=\sum_{\bm z_b} P(\bm z_A, \bm z_B)\\
&=\sum_{\bm z_B} \frac{1-\beta E(\bm z_A, \bm z_B) +\frac{1}{2}\beta^2 E^2(\bm z_A,\bm z_B)}{Z}
+\mathcal{O}\left(\beta^3\right) \ .
\nonumber
\end{align}

Here $\bm z_A$ is fixed and spins $\bm z_B$ in $B$ act like an environment for $A$ and are traced out.

Tracing out the first order term in the numerator of \eqref{PX} annihilates any terms that involve at least one spin in $B$. Therefore, the first order term yields simply the energy of subsystem $A$,
\begin{equation}
E_A(\bm z_A)=-J \sum_{ \langle a , a'  \rangle  \in A}  z_a z_{a'} \ .
\end{equation}
Tracing over the second order term in the numerator of Eq.~(\ref{PX}) involves a double sum over neighbors $\sum_{\langle ij\rangle}\sum_{\langle i'j'\rangle} z_i z_j z_{i'} z_{j'}$. The only combinations of $i,j,i',j'$ that are not annihilated by tracing out are:
\begin{enumerate}
    \item $i,j,i',j'\in A$. Summation over these quadruplets yields $E_A(\bm z_A)^2$.
    \item $i,j,i',j'\in B$. Summation over these quadruplets yields $J^2 N_{{\rm{links}}}^{B}$ where $N_{{\rm{links}}}^{B}$ is the number of links in $B$.
    \item $i\in A, j\in B$ and $\langle i,j\rangle=\langle i',j'\rangle$. Summation over these quadruplets yields $J^2 \ell$ where $ \ell$ is the number of links between $A$ and $B$.
    \item In the triangular lattice there's a fourth option where there exist two distinct spins $i, i' \in A$ which have a common neighbor $j\in B$. The sum over such pairs of spins in $A$ is denoted $\sum_{a a' }'$. 
\end{enumerate}
Therefore, the numerator of \eqref{PX} yields, to second order in $\beta$,
\begin{widetext}
\begin{equation}
P_A(\bm z_a ) =2^{N_B} \frac{1-\beta E_A(\bm z_A)+\frac{1}{2}\beta^2 \left(E_A(\bm z_a)^2 +J^2 N_{{\rm{links}}}^{B}  +J^2  \ell  +J^2  \sum_{a a' }'  z_a z_{a'}\right)  }{Z} +\mathcal{O}\left(\beta^3\right)\ .
\end{equation}

Proceeding with the expansion, plugging in \eqref{Zexpand} and using \protect{$N_{{\rm{links}}}^{ A}+N_{{\rm{links}}}^{ B}+\ell = N_{{\rm{links}}}$}, we get
\begin{align}
P_A(\bm z_a )&=
\frac{1-\beta E_A(\bm z_a)+\frac{1}{2}\beta^2 E_A(\bm z_a)^2 +\frac{1}{2}\beta^2 J^2 \sum_{a a' }'z_a z_{a'}}{Z_A}  +\mathcal{O}\left(\beta^3\right)\ ,\qquad\mbox{with }\label{eq:P} \\
 Z_A&=2^{N^A} \left[1 +\frac{1}{2} \beta^2  J^2 N^A_{\rm{links}} \right] +\mathcal{O}\left(\beta^3\right)\ .
\label{eq:ZA}
\end{align}
\end{widetext}

\eqref{eq:P} has the form of a Boltzmann distribution (note the similarity of \eqref{eq:ZA} to \eqref{Zexpand}) derived from the Hamiltonian $E_A$, with extra couplings generated by the tracing out of $B$ (the last term in the numerator of \eqref{eq:P}). A straightforward but tedious calculation, which will not be detailed here, shows that up to quadratic order in $\beta$ these couplings do not affect the entropy. That is, while they do clearly affect the probabilities of individual states (as explicitly  shown in \eqref{eq:P}) their combined contribution to $S$ cancels out to quadratic order when summed over all states. Therefore, as far as entropy calculations are concerned we can write
\begin{align}
    \begin{split}
    P_A(\bm z_A)&\approx \frac{e^{- \beta E_A(S_a) }}{Z_A}+\mathcal{O}\left(\beta^3\right)\ , \\
    Z_A &= \sum_{\bm z_A} e^{-\beta E_A(\bm z_A)} +\mathcal{O}\left(\beta^3\right)\ ,
    \end{split}
\end{align}
and treat $P_A$ as a standard Boltzmann distribution, for which we have $S=\partial_T(T\log Z)$. Plugging this into \eqref{eq:MI} gives
\begin{align}
\MI(A,B)= \partial_T \left( T \log \frac{Z_{A} Z_{B}}{Z_{X}} \right) \ +\mathcal{O}\left(\beta^3\right)  \ .
\end{align}
Physically the numerator $(Z_{A} Z_{B})$ is the partition function for all the spins in $X$ without the interactions through links connecting $A$ and $B$. Finally, using \eqref{Zexpand} and \eqref{eq:ZA} we obtain the result
\begin{equation}
    \MI_{{\rm{Ising}}}(A,B)=\frac{1}{2} \left( \frac{J}{T}\right)^2 \ell+\mathcal{O}\left(\beta^3\right)\ .
    \label{eq:isinghighT}
\end{equation}

Note that neither the sign of $J$ nor the lattice symmetry (square versus triangular) influence the answer to order $\beta^2$ -- the only relevant parameters are the number of links connecting the two subsystems $\ell$ and the coupling constant $J$. Also, up to this order the magnetic field $H$ does not contribute. A very similar calculation leads to the same form for the XY model, with only a change in the prefactor:
\begin{equation}
    \MI_{{\rm{XY}}}(A,B)=\frac{1}{4} \left( \frac{J}{T}\right)^2 \ell+\mathcal{O}\left(\beta^3\right) \ .
    \label{eq:xyhighT}
\end{equation}
Both \eqref{eq:isinghighT} and \eqref{eq:xyhighT} are valid also when $A$ and $B$ do not compose the whole system, but are a part of a larger system.

%%% Add this line AFTER all your figures and tables
%\FloatBarrier

\subsection{Low-temperature expansion - XY model in a magnetic field}
Statistical mechanics problems of continuous variables can be treated at low temperatures via an harmonic treatment of the interactions, i.e.~a mapping to a system of coupled harmonic oscillators. This technique can be applied to compute $\MI$ too~\cite{katsinis2020inverse}, yielding closed-form formulas. Here we apply this method to the XY model in an external magnetic field ($H$) in the zero-temperature limit.

The XY model in a magnetic field is defined by the partition function
\begin{align}
\begin{split}
Z &= \int_0^{2 \pi } d \bm \theta e^{- \beta E(\bm\theta)},  \\
E(\bm\theta)&=-J \sum_{\langle i,j \rangle}\cos(\theta_i - \theta_j)-H \sum_i \cos \theta_i.
\end{split}
\end{align}

At low temperature $T \ll J,H$ the variables $\bm\theta$ explore only the vicinity of the minimum of the external potential $-H \cos \theta_i$, and since we consider a frustration-free lattice (square lattice), also the differences $\theta_i - \theta_j$ on neighbouring links $\langle i,j \rangle$ will be located near the minima of $-J \cos(\theta_i - \theta_j)$. Performing a harmonic approximation of the overall potential we get:
\begin{align}
\begin{split}
Z_0 &=  \int_{-\infty}^{\infty} d \bm \theta e^{- \beta E_0(\bm \theta)}\ , \\
E_0(\bm\theta)&=\frac{J}{2} \sum_{\langle i,j \rangle}(\theta_i - \theta_j)^2  + \frac{H}{2} \sum_i \theta_i^2 + {\rm{const}}\ .
\end{split}
\end{align}
Here, we extended the variables $\theta_i$ from being angles to unconstrained real numbers. Accordingly, microstates of the full system $X$ satisfy a multivariate normal distribution 
\begin{align}\label{eq:Pnorm}
\begin{split}
p(\bm \theta) &= \frac{e^{-\frac{1}{2} \bm\theta^T M \bm\theta }}{Z_0}\ ,
\qquad\mbox{with}\\
M_{ij}&=\frac{H+zJ}{T}\delta_{ij} -\frac{J}{T}\delta_{\langle i,j \rangle} \ .
\end{split}
\end{align}
$M$ is the system's Hessian, a $N \times N$ matrix where $N$ is the number of sites in the system $X$. Here $z$ is the coordination number ($z=4$ for a square lattice) and $\delta_{\langle i,j \rangle}=1$ if $i$ and $j$ are neighbors and 0 otherwise. The entropy of a multivariate Gaussian is well known:
\begin{align}\label{eq:entropyM}
S(X)=\frac{N}{2} \log 2 \pi  e - \frac{1}{2} \log \det M.
\end{align}
For a single spin in a magnetic field, for example, this gives $S=\log \left(\sqrt{2 \pi e T/H}\right)$ which is valid as long as the variance of $\theta$, $(T/H)^2$, is sufficiently small compared to $(2 \pi)^2$. 

The key object required for the calculation of the $\MI$ is the marginal probability for a subsystem $A$.  It is obtained  by integrating $p(\bm \theta)$ over all degrees of freedom $\bm\theta_B\in B$,
\begin{align}
p_A(\bm\theta_A) =\frac{1}{Z}\int_0^{2 \pi } d \bm \theta_B e^{- \beta E(\bm\theta_A,\bm \theta_B)}.
\end{align}
To perform the Gaussian integral we decompose the matrix $M$ as
\begin{align}\label{MAB}
 M = \begin{pmatrix}
    M_{AA} & M_{AB}\\\
    M_{BA} & M_{BB}
\end{pmatrix},
\end{align}
where, $M_{AA}$ is an $N^A \times N^A$ matrix acting only on the $N^A$ degrees of freedom in $A$, and similarly for $M_{BB}$. The off-diagonal blocks $M_{AB}=M_{BA}^T$ couple the two subsystems. Thus,
\begin{widetext}
\begin{align}\label{eqffcovariance}
p_A( \bm\theta_A ) &=e^{-\frac{1}{2} \bm\theta_A^T M_{AA} \bm\theta_A} \int d\theta_B 
\exp\left[-\frac{1}{2}\bm\theta_B^T M_{BB} \bm\theta_B -  \bm\theta_A^T M_{AB}\bm\theta_B\right] \ .
\end{align}

Performing the Gaussian integral over $\bm\theta_B$ gives
\begin{align}
P(\bm \theta_A) =
\left((2 \pi)^{N_B} \det M_{BB}\right)^{1/2}
 \exp\left[
 -\frac{1}{2} \bm\theta_A^T M_{AA} \bm\theta_A
 -\frac{1}{2} \bm\theta_A^T \left(M_{AB}   M_{BB}^{-1} M_{BA} \right) \bm\theta_A
 \right] \ .
\end{align}
\end{widetext}
Since the marginal distribution is also Gaussian, its entropy is given by \eqref{eq:entropyM}, with the effective Hessian (covariance matrix) of $A$ given by \eqref{eqffcovariance},
\begin{align}
M_{A}^ {\rm{eff}} = M_{AA}-  M_{AB}   M_{BB}^{-1} M_{BA}\ .
\label{meff}
\end{align}
$M_{A}^ {\rm{eff}}$ contains direct interactions inside $A$, as well as new interactions $M_{AB}   M_{BB}^{-1} M_{BA}$ generated by tracing out the environment $B$. We thus have
\begin{equation}
\label{MIdet}
\MI = \frac{1}{2} \log\frac{ \det M_{X}}{\det M_{A}^{\rm{eff}}\det M_{B}^{\rm{eff}}} .
\end{equation}

Note that this expression gives the $T\to0$ limit of $\MI$ and is independent of $T$. Finite temperature corrections are not present in the harmonic approximation and start to appear when the variance of spins becomes of order $2 \pi$ and deviations from the Gaussian distribution are sampled.

For the system described in the main text $\MI$ was computed by evaluating the determinant in \eqref{eq:entropyM} numerically using the effective covariance matrix \eqref{meff}.

%%%%%%%%%%%%%%%%%%%%%%%%%%%%%%%%%%%%%%%%%%%%%%%%%%%%%%%%%%%%%%%%%%%%%%%%%%%%
\section{$\MI$ between two XY-spins in a magnetic field} 
\begin{figure*}
\centering
\includegraphics[width=\linewidth]{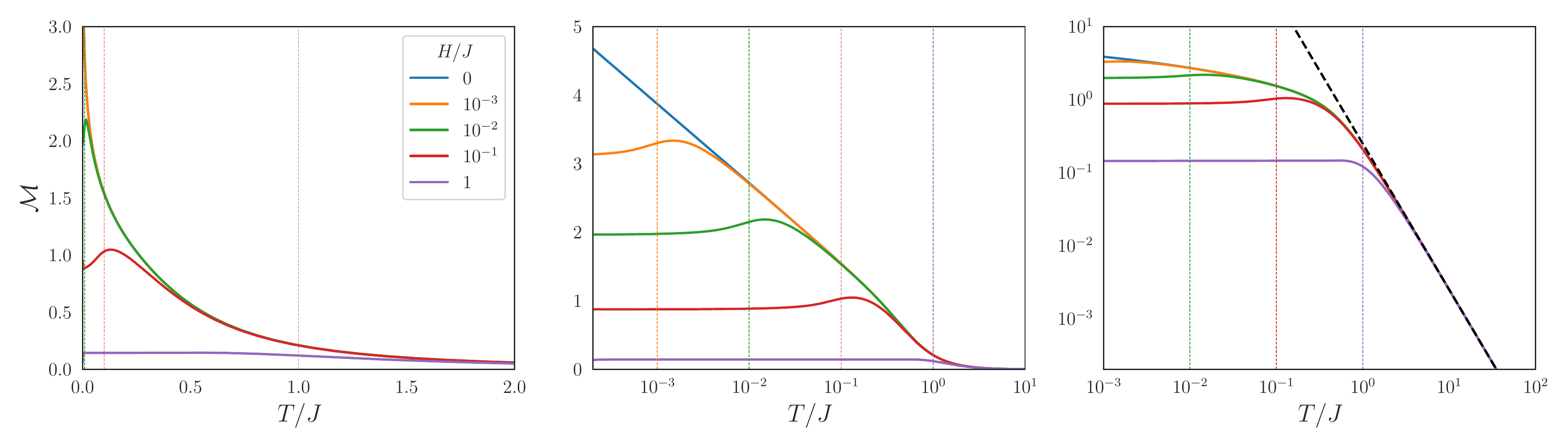}
\caption{Exact calculation of $\MI$ for two XY spins ($J=1$) in the presence of external field ($H$). The same data are shown in linear-linear, log-linear and log-log scales (some data of the middle panel appears also in the main text). Colored vertical dashed lines show $T=H$ with the color code corresponding to $H$ as in the legend. The dashed black line in the right panel is the high temperature expansion limit of \eqref{eq:xyhighT}.}
\label{fig:exact2spins}
\end{figure*}

It is instructive to contrast the result in the main text for the $\MI$ of the $XY$ model with that for a system consisting of only two spins. This can be calculated exactly, and is shown in Fig.~\ref{fig:exact2spins}. At high temperature $\MI$ decreases like $\MI \to \frac{1}{4} \left(\frac{J}{T} \right)^2$, indicated by a dashed line in the right panel, as predicted by \eqref{eq:xyhighT}. As $T \to 0$, we can see in the central panel a logarithmic divergence with $T$ which is cut-of when $T\approx H$. 

Indeed it is easy to derive from Eqs.~(\ref{eq:Pnorm}), (\ref{meff}) and (\ref{MIdet}) the zero temperature limit of $\MI$,
\begin{align}
\lim\limits_{T \to 0}\MI_{\rm{two~spins}} = \log \frac{H+J}{\sqrt{H(H+J)}}.
\end{align}
As $H$ increases, the cutoff of the logarithmic divergence occurs at higher temperatures, and the peak thus shifts to higher temperatures. Thus,the peak itself, as well as its $H$-dependence features, are already present in a two-spin system. 

%%%%%%%%%%%%%%%%%%%%%%%%%%%%%%%%%%%%%%%%%%%%%%%%%%%%%%%%%%%%%%%%%%%%%%%%%%%%
\section{Simulations of the soft sphere system}
The system is an equimolar system of larger and smaller spheres. We choose the units such that the diameter of the smaller sphere is unity, and the radius of the larger one is 1.4. The dynamics were simulated using a fast inertial relaxation engine algorithm~\cite{PhysRevLett.97.170201} in a square box of size 150 with periodic boundary conditions. 100 realizations were generated for each $\phi$, ranging between 14,000 to 17,000 particles.

%%%%%%%%%%%%%%%%%%%%%%%%%%%%%%%%%%%%%%%%%%%%%%%%%%%%%%%%%%%%%%%%%%%%%%%%%%%%
\section{Discrete vs. differential entropy}
As discussed in the main text, the system of bidisprese sphere is a continuous system, parameterized by a continuous vector $\bm x\in\mathbb{R}^{2N}$ where $N$ is the number of particles in the system. However, the state of the system is provided to the ANN as a binary image, which is a discrete variable. Here we discuss the subtleties of comparing the discrete and continuous defintions of entropy (Eq.~(1) and (7) of the main text, respectively).

Let us denote $p(\bm x)$ the probability density of observing the configuration $\bm x$. The discretization is a mapping of the continuous vector $\bm x$ to an image $I(\bm x)$ where $I$ takes one of a finite set of values which we denote $I_1, I_2, \dots$. Each $I_i$ is associated with its pre-image $\Omega_i$, observation probability $p_i$ and phase-space volume $v_i$, defined as follows:
\begin{align}
    \Omega_i & \equiv \left\{\bm x\ |\ I(\bm x)=I_i\right\} \ ,\nonumber &
    p_i &\equiv \int_{\Omega_i} p(\bm x) d\bm x\ , \ 
    \\
    v_i &\equiv \int_{\Omega_i} 1\, d\bm x\ . 
    \label{eq:discretization}
\end{align}
In the limit of very fine discretization, i.e.~$\max_i\{v_i\}\to 0$, and assuming $p(x)$ is not ill-behaved, the second definition can be approximated as
\begin{align}
    p_i \approx p(\bm x_i) v_i \ ,
    \label{eq:approx_discretization}
\end{align}
where $\bm x_i$ is any point in $\Omega_i$. This approximation is accurate when the discretization is fine enough such that $p$ doesn't change considerably across $\Omega_i$, i.e.~when all configurations that are mapped to the same image are roughly equiprobable. When this happens, the differential entropy $\tilde S$ can be approximated by a Riemman sum:
\begin{widetext}
\begin{equation}
\begin{split}
    \tilde S 
    &= -\int p(\bm x) \log p(\bm x)d\bm x 
    \approx-\sum_i  \Big(p(\bm x_i) \log p(\bm x_i) \Big)\cdot v_i\\
    &\approx-\sum_i\left(\frac{p_i}{v_i} \log \left(\frac{p_i}{v_i}\right) \right)\cdot v_i
    = \sum_i\left(- p_i \log p_i + p_i\log v_i \right) = S+\sum_i p_i\log v_i \ .
\end{split}
\label{eq:continuum_vs_discrete}
\end{equation}
We see that $\tilde S$ differs from $S$ by a term logarithmic in the resolution size. This term, however, cancels out when computing $\mathcal M$ rather than $S$.

To see this, let's say $\bm x$ and $\bm y$ are random variables, with the joint probability density $p(\bm x, \bm y)$ and marginal densities $p^x(\bm x)=\int p(\bm x, \bm y)d\bm y$ and $p^y(\bm y)=\int p(\bm x, \bm y)d\bm x$. In addition, we have two discretization schemes $I^x(\bm x)$ and $I^y(\bm y)$ that map each observation to some finite set. We define, in analogy to \eqref{eq:discretization}, 
\begin{align*}
    \Omega_{ij} & \equiv \left\{(\bm x, \bm y)\ |\ I^x(\bm x)=I^x_i\mbox{ and }I^y(\bm x)=I^y_j \right\} \ ,&
    p_{ij} &\equiv \int_{\Omega_{ij}} p(\bm x, \bm y) d\bm x\ , \ &
    v_{ij} &\equiv \int_{\Omega_{ij}} 1\, d\bm xd\bm y\ , \\
    \Omega^x_i & \equiv \left\{\bm x\ |\ I^x(\bm x)=I^x_i\right\} \ ,&
    p^x_i &\equiv \int_{\Omega^x_i} p^x(\bm x) d\bm x\ , \ &
    v^x_i &\equiv \int_{\Omega^x_i} 1\, d\bm x\ , \\
    \Omega^y_j & \equiv \left\{\bm y\ |\ I^y(\bm y)=I^y_j\right\} \ ,&
    p^y_j &\equiv \int_{\Omega^y_j} p^y(\bm y) d\bm y\ , \ &
    v^y_j &\equiv \int_{\Omega^y_j} 1\, d\bm y\ . \\
\end{align*}
\end{widetext}
Eqs.~(1)-(2) of the main text can be combined to represent the mutual information as
\begin{align}
\mathcal{M} &= \int p(\bm x, \bm y)\log\left(\frac{p(\bm x, \bm y)}{p^x(\bm x)p^y(\bm y)}\right)d\bm x\,d\bm y
\label{eq:M}
\end{align}
Since $v_{ij}=v^x_iv^y_j$, the analog of \eqref{eq:approx_discretization} is
\begin{align*}
    p_{ij} &\approx p(\bm x_i, \bm y_j) v^x_iv^y_j \ , &
    p^x_i &\approx p^x(\bm x_i) v^x_i \ , &
    p^y_j &\approx p^y(\bm y_j) v^y_j \ . &
\end{align*}
Finally, combining all the above we get
\begin{align}
\begin{split}
\mathcal{M} &
\approx \sum_{i,j} p(\bm x_i, \bm y_j)\log\left(\frac{p(\bm x_i, \bm y_j)}{p^x(\bm x_i)p^y(\bm y_j)}\right)v^x_i v^y_j \\
&\approx \sum_{i,j} p_{ij}\log\left(\frac{p_{ij}}{p^x_i p^y_j}\right)\ ,
\end{split}
\end{align}
which identifies with the discrete defintion of $\mathcal M$.

As an aside, we note that \eqref{eq:continuum_vs_discrete} has an intuitive interpretation: $\log v_i$ is exactly the entropy of a uniform distribution over $\Omega_i$ (whose probability density is $p=1/v_i$). Therefore, the differential entropy $\tilde S$ measures the uncertainty (=entropy) associated with knowing in which $\Omega_i$ the observation $\bm x$ resides, plus the average uncertainty (=entropy) associated with knowing where does $\bm x_i$ resides within $\Omega_i$. The latter cancels out when computing $\mathcal{M}$.

%%%%%%%%%%%%%%%%%%%%%%%%%%%%%%%%%%%%%%%%%%%%%%%%%%%%%%%%%%%%%%%%%%%%%%%%%%%%
\section{Derivation of Eq.~(10) of the main text}
Eq.~(4) of the main text starts with a system $X_0$ of a given volume $V_0$ and looks at smaller and smaller subsystems (i.e.~larger $m$). For the purposes of Eq.~(10) of the main text we want to explore the other direction -- assuming that $X_0$ is by itself a part of a much larger system and extrapolating from $X_0$ to the system size. To comply with the notation of the main text, where larger $m$'s correspond to smaller subsystems $X_m$, we consider subsystems which  are formally indexed by negative integers. Also, it will be useful to consider Eq.~(3) of the main text normalized per unit volume.  For  any $k$ we have
\begin{align*}
S(X_{k-1}) &= 2S(X_k)-\MI(X_k) 
\qquad\Rightarrow\\
s(X_{k-1}) &\equiv\frac{S(X_{k-1})}{V_{k-1}}= s(X_{k})-\frac{\MI(X_{k})}{2V_{k}}\  ,
\end{align*}
where we used the fact that $V_{k-1}=2V_k$. Using this relation recursively we get
\begin{widetext}
\begin{equation}
\begin{split}
s(X_{-1}) &= s(X_0)-\frac{\MI(X_{0})}{2V_{0}}\\
s(X_{-2}) =s(X_{-1})-\frac{\MI(X_{-1})}{2V_{-1}} &= s(X_0)-\frac{\MI(X_{0})}{2V_{0}}-\frac{\MI(X_{-1})}{4V_{0}}\\
s(X_{-3}) =s(X_{-2})-\frac{\MI(X_{-2})}{2V_{-2}} &= s(X_0)-\frac{\MI(X_{0})}{2V_{0}}-\frac{\MI(X_{-1})}{4V_{0}}-\frac{\MI(X_{-2})}{8V_{0}}\\
&\vdots\\
s(X_{-m}) &= s(X_0)-\frac{1}{2V_0}\sum_{k=0}^{m-1} \frac{\MI(X_{-k})}{2^{k}}\\
\end{split}
\end{equation}
\end{widetext}
We now assume that for subsystems larger than $X_0$ the mutual information is extensive, so by Eq.~(9) of the main text we have $\MI(X_{-k}) = (\ell_{-k}/\ell_0 )\MI(X_0)$. For our choice of selecting subsystems, we also have $\ell_{-k}/\ell_0=2^{\left\lfloor\frac{k+1}{2}\right\rfloor}$, where $\left\lfloor\cdot\right\rfloor$ is the floor function. We assume that $X_0$ is a square subsystem (subsystems alternate between square and rectangular, cf.~Fig.~1 of the main text). Putting all this together we get
\begin{align}
  S(X_{-m}) &= s(X_0)-\frac{\MI_0}{2V_0}\sum_{k=0}^{m-1} 2^{\left\lfloor\frac{k+1}{2}\right\rfloor-k}\ .
\end{align}
One can easily verify that in the limit $m\to\infty$ the sum in the last equation approaches 4. We conclude that
\begin{align}
  s(X_{-m}) &= s(X_0)-2\frac{\MI_0}{V_0}\ .
\end{align}

\begin{figure*}
\centering
\includegraphics[width=0.7\linewidth]{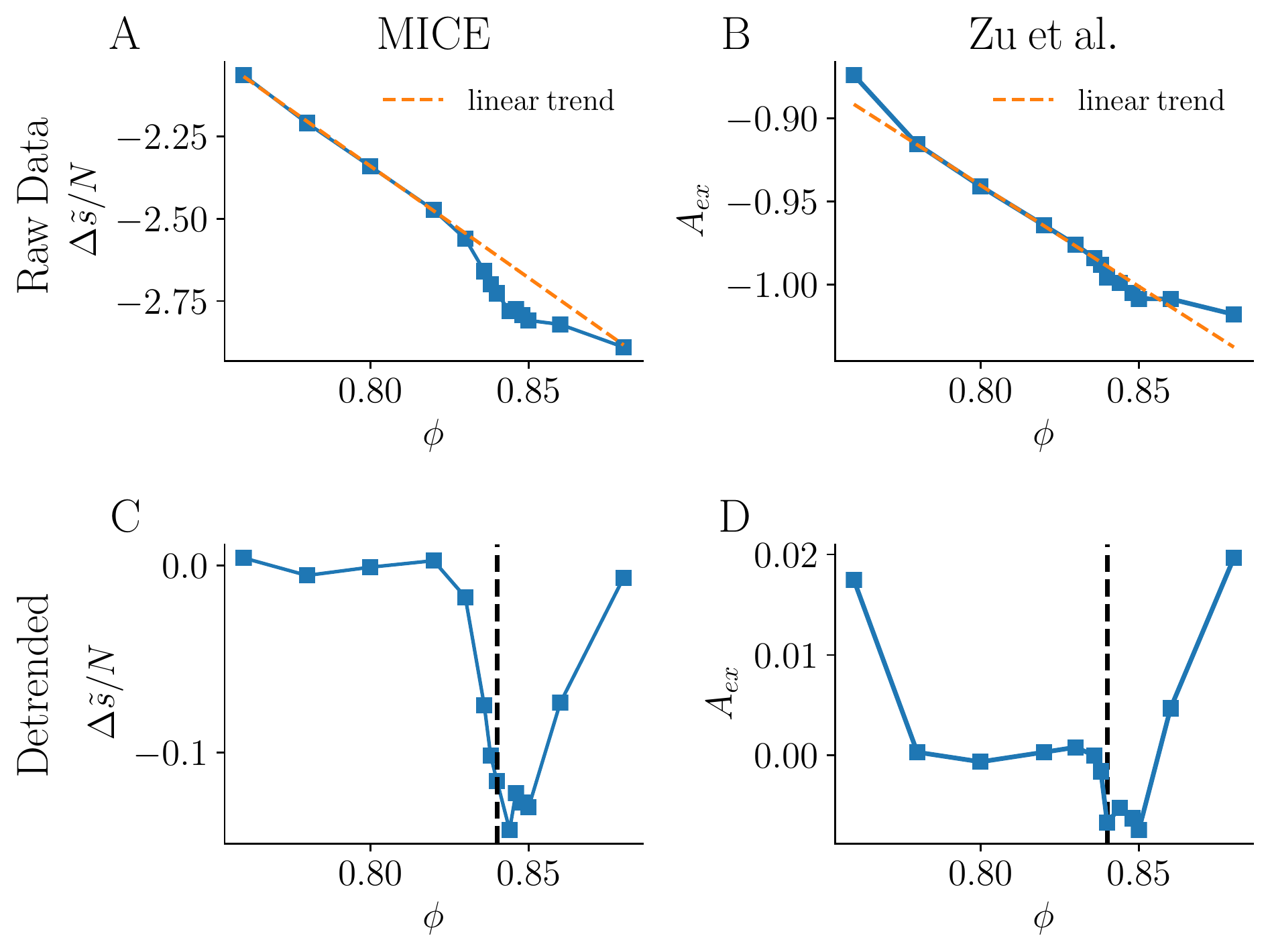}
\caption{Entropy estimation of the bidisperse soft sphere mixture, using two different methods, see text of Sec.~\ref{sec:compare} for a description.  The dashed black line represents the theoretical jamming transition point.}
\label{fig:CompareCompression}
\end{figure*}

%%%%%%%%%%%%%%%%%%%%%%%%%%%%%%%%%%%%%%%%%%%%%%%%%%%%%%%%%%%%%%%%%%%%%%%%%%%%
\section{Comparing \emph{MICE} and the results of Zu et.~al.}
\label{sec:compare}
In the main text we claimed that \emph{MICE} outperforms the compression method used by Zu et.~al.~\cite{Daan} in detecting the jamming point. This was based on their statements that their Computable Information Density (CID) estimates do not show a minimum near the jamming point (see Sec.~3.5 of their paper).

In Fig.~\ref{fig:CompareCompression} we show a direct comparison between our data (left column, the same data appear as Fig.~4F and its inset in the main text) and theirs (right column, taken from Figure 6A of~\cite{Daan}). 

The top row shows the estimation of the ``excess entropy'', i.e.~the difference in entropy from some baseline behavior: With \emph{MICE} this is achieved by omitting the entropic contribution of the smallest scales, cf.~Eq.~(8) of the main text. Zu et.~al.~do this by subtracting the information density of an ideal gas (see Sec.~2.3.2 of Zu et.~al.~\cite{Daan}). These two baselines are conceptually similar but quantitatively different and therefore the absolute numbers differ somewhat between the methods. The trend, however, is informative.

To better visualize the signature of the transition, in the bottom row we plot the same data as the top row, with a linear trend (shown in dashed orange in the top row) subtracted. It is seen that the deviations from linearity are very pronounced when measured with \emph{MICE}, but the CID estimation shows small deviations compared to the overall effect.

\end{document}